\documentclass[a4paper,11pt]{article}

\usepackage{jcappub} 
\usepackage{amssymb}
\usepackage{amsthm}

\usepackage{graphicx}

\usepackage[T1]{fontenc} 

\title{The wavefront of the radio signal emitted by cosmic ray air showers}

\author[a]{W.D.~Apel}
\author[b]{J.C.~Arteaga-Vel\'azquez}
\author[c]{L.~B\"ahren}
\author[a]{K.~Bekk}
\author[d]{M.~Bertaina}
\author[e]{P.L.~Biermann}
\author[a,f]{J.~Bl\"umer}
\author[a]{H.~Bozdog}
\author[g]{I.M.~Brancus}
\author[d,h]{E.~Cantoni}
\author[d]{A.~Chiavassa}
\author[a]{K.~Daumiller}
\author[i]{V.~de~Souza}
\author[d]{F.~Di~Pierro}
\author[a]{P.~Doll}
\author[a]{R.~Engel}
\author[c,e,j]{H.~Falcke}
\author[f]{B.~Fuchs}
\author[l]{H.~Gemmeke}
\author[m]{C.~Grupen}
\author[a]{A.~Haungs}
\author[a]{D.~Heck}
\author[j]{J.R.~H\"orandel}
\author[e]{A.~Horneffer}
\author[f]{D.~Huber}
\author[a]{T.~Huege}
\author[n]{P.G.~Isar}
\author[k]{K.-H.~Kampert}
\author[f]{D.~Kang}
\author[l]{O.~Kr\"omer}
\author[j]{J.~Kuijpers}
\author[f]{K.~Link}
\author[o]{P.~{\L}uczak}
\author[f]{M.~Ludwig}
\author[a]{H.J.~Mathes}
\author[f]{M.~Melissas}
\author[h]{C.~Morello}
\author[a]{J.~Oehlschl\"ager}
\author[f]{N.~Palmieri}
\author[a]{T.~Pierog}
\author[k]{J.~Rautenberg}
\author[a]{H.~Rebel}
\author[a]{M.~Roth}
\author[l]{C.~R\"uhle}
\author[g]{A.~Saftoiu}
\author[a]{H.~Schieler}
\author[l]{A.~Schmidt}
\author[a]{S.~Schoo}
\author[a,1]{F.G.~Schr\"oder,\note{Corresponding author.}}
\author[p]{O.~Sima}
\author[g]{G.~Toma}
\author[h]{G.C.~Trinchero}
\author[a]{A.~Weindl}
\author[a]{J.~Wochele}
\author[o]{J.~Zabierowski}
\author[e]{J.A.~Zensus}

\affiliation[a]{Institut f\"ur Kernphysik, Karlsruhe Institute of Technology (KIT), Germany}
\affiliation[b]{Universidad Michoacana, Morelia, Mexico}
\affiliation[c]{ASTRON, Dwingeloo, The Netherlands}
\affiliation[d]{Dipartimento di Fisica Generale dell' Universit\`a Torino, Italy}
\affiliation[e]{Max-Planck-Institut f\"ur Radioastronomie Bonn, Germany}
\affiliation[f]{Institut f\"ur Experimentelle Kernphysik, Karlsruhe Institute of Technology (KIT), Germany}
\affiliation[g]{National Institute of Physics and Nuclear Engineering, Bucharest, Romania}
\affiliation[h]{INAF Torino, Instituto di Fisica dello Spazio Interplanetario, Italy}
\affiliation[i]{Universidad S\~ao Paulo, Inst. de F\'{\i}sica de S\~ao Carlos, Brasil}
\affiliation[j]{Radboud University Nijmegen, Department of Astrophysics, The Netherlands}
\affiliation[k]{Universit\"at Wuppertal, Fachbereich Physik, Germany}
\affiliation[l]{Institut f\"ur Prozessdatenverarbeitung und Elektronik, Karlsruhe Institute of Technology (KIT), Germany}
\affiliation[m]{Universit\"at Siegen, Fachbereich Physik, Germany}
\affiliation[n]{Institute for Space Sciences, Bucharest, Romania}
\affiliation[o]{National Centre for Nuclear Research, Department of Astrophysics, {\L}\'{o}d\'{z}, Poland}
\affiliation[p]{University of Bucharest, Department of Physics, Romania}

\emailAdd{frank.schroeder@kit.edu}

\abstract{Analyzing measurements of the LOPES antenna array together with corresponding CoREAS simulations for more than 300 measured events with energy above $10^{17}\,$eV and zenith angles smaller than $45^\circ$, we find that the radio wavefront of cosmic-ray air showers is of approximately hyperbolic shape. The simulations predict a slightly steeper wavefront towards East than towards West, but this asymmetry is negligible against the measurement uncertainties of LOPES. At axis distances $\gtrsim 50\,$m, the wavefront can be approximated by a simple cone. 
According to the simulations, the cone angle is clearly correlated with the shower maximum. Thus, we confirm earlier predictions that arrival time measurements can be used to study the longitudinal shower development, but now using a realistic wavefront. 
Moreover, we show that the hyperbolic wavefront is compatible with our measurement, and we present several experimental indications that the cone angle is indeed sensitive to the shower development. Consequently, the wavefront can be used to statistically study the primary composition of ultra-high energy cosmic rays. At LOPES, the experimentally achieved precision for the shower maximum is limited by measurement uncertainties to approximately $140\,$g/cm\textsuperscript{2}. But the simulations indicate that under better conditions this method might yield an accuracy for the atmospheric depth of the shower maximum, $X_\mathrm{max}$, better than $30\,$g/cm\textsuperscript{2}. This would be competitive with the established air-fluorescence and air-Cherenkov techniques, where the radio technique offers the advantage of a significantly higher duty-cycle. Finally, the hyperbolic wavefront can be used to reconstruct the shower geometry more accurately, which potentially allows a better reconstruction of all other 
shower parameters, too.}


\begin{document}
\maketitle
\flushbottom


\section{Introduction}
\label{sec_introduction}

At the current state of technical development, air showers are the only access to ultra-high-energy cosmic-ray physics. Radio measurements of air showers become effective at primary particle energies $\gtrsim 10^{17}\,$eV. Distinguishing different scenarios for the still unknown origin of ultra-high energy cosmic rays requires knowledge of the cosmic-ray composition at these energies.
Measuring the longitudinal air-shower development is the presently best method for the reconstruction of the mass of the primary particles, since the shower development depends on a statistical basis on the mass of the primary particle: heavy nuclei interact on average earlier in the atmosphere than light nuclei. 
Thus, the air-shower development of heavy nuclei starts at a higher altitude, which leads to a lower atmospheric depth of the shower maximum, $X_\mathrm{max}$, i.e.~a larger distance to ground, and consequently causes different signatures in all secondary products (particles and radiation) of the air shower.

Established methods for $X_\mathrm{max}$ measurements and the derived cosmic ray composition are the detection of air-fluorescence and air-Cherenkov light of air showers. However, these methods are restricted to dark nights and good weather conditions.
The radio emission of air showers is also sensitive to the longitudinal shower development \cite{2012ApelLOPES_MTD}, and can be detected with a considerably higher duty-cycle of almost $100\,\%$.
However, it still has to be demonstrated that the $X_\mathrm{max}$ precision of radio measurements can be competitive, not only for very dense antenna arrays like LOFAR \cite{BuitinkLOFARIcrc2013}, but also for sparser, large-scale arrays which can be built at reasonable costs.

The origin of the radio emission of air showers is well described in other papers. For instance, recent overviews are available in Refs.~\cite{RevenuExperimentsOverview_ARENA2012, HuegeTheoryOverview_ARENA2012, HuegeIcrc2013}. At typical measurement distances up-to a few $100\,$m of the shower axis, the radio emission is coherent for frequencies in the order of $\lesssim 100\,$MHz. 
The dominant mechanism for the emission is the geomagnetic deflection of the secondary electrons and positrons in the air shower inducing time-varying transverse currents \cite{KahnLerche1966, FalckeGorham2003}, but also other effects play a role. In particular, the Askaryan effect \cite{Askaryan1962, AugerAERApolarization2014} is non-negligible, i.e., radio emission due to a time variation of the net charge in the shower. 
Moreover, the refractive index of the air influences the coherence conditions, which leads to observable effects in the lateral distribution of the radio amplitude \cite{Werner2012}, and also affects the radio wavefront. All these effects are included in the CoREAS \cite{HuegeCoREAS_ARENA2012} Monte-Carlo-simulation code used for this paper. Other contributions to the radio emission might exist, but have not yet been predicted or demonstrated to be significant.

The results presented in this paper, are based on interferometric LOPES measurements and CoREAS simulations made for the measured events. LOPES is a first-generation digital antenna array which was operating from 2003 to 2013 at the Karlsruhe Institute of Technology (KIT), Germany.
LOPES is limited in precision, in particular because of the limited size and the large radio background. Nevertheless, LOPES has the advantage that it is triggered by the co-located particle-detector array of the KASCADE-Grande experiment, which provides well-calibrated measurements of the same air shower.
Consequently, LOPES is a pathfinder experiment to study the principles of the radio emission and to develop analysis techniques for the radio measurements. But, it is not a precision experiment, which already could compete with the precision of established techniques for air showers, like air-fluorescence, air-Cherenkov, or secondary particle measurements. Consequently, the focus of the present analysis is the principal dependencies of the radio wavefront on other shower parameters and the development of a specific method for $X_\mathrm{max}$ reconstruction. More detailed analyses aiming at higher accuracy can better be done by recently started antenna arrays, like LOFAR \cite{SchellartLOFAR2013}, AERA \cite{Melissas_ARENA2012} and Tunka-Rex \cite{TunkaRexRICAP2013}.

The longitudinal shower development and, thus, $X_\mathrm{max}$ can be accessed from radio measurements in at least three different ways:

First, the lateral distribution of the radio amplitude \cite{HuegeUlrichEngel2008, deVries2010, 2012ApelLOPES_MTD}. For showers with more distant shower maxima, the lateral distribution is flatter.
Vice versa, for closer shower maxima the amplitude decreases faster with increasing distance to the shower axis. This method has recently been exploited for a first reconstruction of $X_\mathrm{max}$ based on measured LOPES events \cite{2014ApelLOPES_MassComposition}.

Second, the slope of the frequency spectrum of the radio signal and the pulse shape depend not only on the distance to the shower axis, but also contain information on $X_\mathrm{max}$ \cite{Grebe_ARENA2012}. However, the applicability and achievable precision of this method has not yet been studied in detail. At LOPES, this method cannot be applied due to the strong radio background.

Third, the shape of the radio wavefront depends on the longitudinal shower development. This has been theoretically studied earlier \cite{Lafebre2010}, and the present work brings significant news in several aspects. 
Newer simulations are used including all known, significant effects of the radio emission, and a new analytical description, namely the hyperbolic wavefront, is proposed. 
Moreover, for the first time, the wavefront is applied to measured data for the reconstruction of the shower maximum. In addition, the hyperbolic wavefront brings the benefit that the reconstructed shower geometry can be improved compared to spherical or plane wave assumptions.

\begin{figure}[t]
  \centering
  \includegraphics[width=0.7\columnwidth]{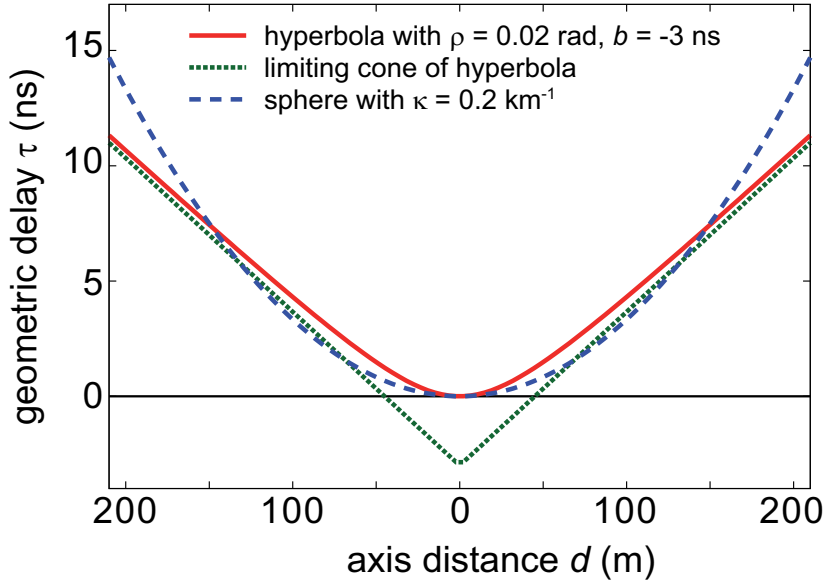}
  \caption{Comparison of a spherical, a conical, and a hyperbolic wavefront for typical parameters. While the spherical wavefront is a good approximation for the hyperbola close to the shower axis, the conical wavefront becomes a sufficient approximation at axis distances $d \gtrsim 50\,$m. The parameters $\rho$, $b$, and $\kappa$ are defined later by the equations in section \ref{section_AnalyticDescirption}.}
   \label{fig_wavefront_examples}
\end{figure}

\section{Definition of the Radio Wavefront}

There are several ways to define the radio wavefront, and certainly several ways to determine it from data. We define the radio wavefront as the surface where the radio amplitude is maximum.
Obviously this definition implies that the radio wavefront depends on the observation level and also on the observed frequency band, since this determines the pulse shape. Consequently, our analysis is in particular valid for the effective bandwidth of LOPES which is $43-74\,$MHz and the altitude of LOPES, which is $110\,$m above see level. The results and the parameters used for the reconstruction of $X_\mathrm{max}$ might be different for other frequency bands and observation levels.
In particular, the bandwidth affects the measured pulse shape and, consequently, any method to determine the arrival time of the signal --- be it the time of the maximum used for the CoREAS simulations, or the cross-correlation beamforming used for the LOPES measurements.

Furthermore, the wavefront propagates towards Earth with the air shower and potentially changes its shape during shower development. Therefore, we also have to define a time, at which we study the wavefront. The natural choice is the time $t_0$ when the wavefront touches ground at the air shower axis.
$t_0$ is known for the simulations. In principle, $t_0$ is accessible also by experiments, but practically a measurement is challenging, and the consequences for LOPES will be discusses later.

While for the radio amplitude asymmetries with respect to the shower axis have been observed \cite{CODALEMAMarinICRC2011, NellesLOFAR_LDF2014}, and are expected due to the interference of the geomagnetic and the Askaryan effect during the radio emission,the symmetry of the wavefront had not been studied, yet. We find a small asymmetry of the wavefront in the CoREAS simulations: For the geomagnetic field at the LOPES site, the wavefront is slightly steeper towards East than towards West. However, the size of the asymmetry is negligible compared to our measurement precision (for details see section \ref{sec_assymetry}).

For this reason, we studied only radio wavefronts symmetric around the shower axis. 
In the simplest approximation the radio wavefront is a plane perpendicular to the shower axis, which corresponds to an infinite distant source.
This approximation is sufficient to estimate the shower direction within a few degrees, but already early LOPES results \cite{NiglDirection2008} showed that this is oversimplified compared to the achieved measurement precision.

The next reasonable approximations are wavefronts depending on one free parameter, in particular a sphere and a cone. While a sphere would correspond to a static point source approximation, a moving source or a line source ought to generate a conical wavefront - like a ship on a lake generates a conical bow wave.
As we will see, a combination of a sphere and a cone, namely a hyperbola, gives an even better description for the radio wavefront of air showers:
It is spherical very close to the shower axis, which might be an effect of the finite extension of the air shower, and it approximates a cone at larger distances.

Figure \ref{fig_wavefront_examples} shows these wavefronts using parameters corresponding to a typical near vertical shower measured with LOPES.
The figure illustrates that for an antenna array like LOPES with an extension of $200\,$m and a typical distance between antennas of $\sim 30\,$m, it is difficult to distinguish between the different non-planar wavefronts: the typical event has no or only a few antennas in the relevant distance range, very close or very distant from the shower axis.
Still, the simulations clearly point to a hyperbolic wavefront, and with LOPES we achieve the highest precision for the arrival direction using the hyperbolic model. Moreover, recent measurements by LOFAR support that the wavefront is of approximately hyperbolic shape \cite{CorstanjeLOFAR_wavefront2014}.

\begin{figure*}[t]
  \centering
  \includegraphics[width=0.7\linewidth]{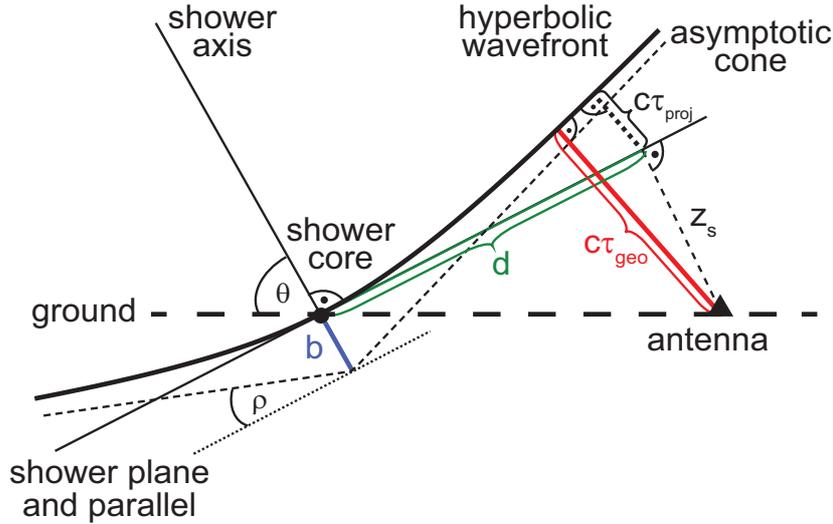}
  \caption{Geometrical delays $\tau_\mathrm{geo}(d,z_s)$ in dependence of the antenna position in shower coordinates $(d,z_s)$ for a hyperbolic wavefront; $b$ is the offset parameter and $\rho$ the cone angle of the asymptotic cone which is approached by the hyperbola at large distances; $c$ is the speed of light, $\theta$ the zenith angle of the air shower, and $\tau_\mathrm{proj}(d)$ is the geometric delay when the antenna position is projected to the shower plane ($d, z_s, b$ are parameters in space [m] or time [ns], respectively, if multiplied with $c$; $\rho, \theta$ are angles [rad]).}
   \label{fig_hyperbola}
\end{figure*}

\subsection{Analytic description}
\label{section_AnalyticDescirption}
The radio wavefront can be analytically described as function of the geometric delay $\tau_\mathrm{geo}$ at each antenna position, which is the delay of the wavefront with respect to the shower plane, i.e., the plane perpendicular to the shower axis containing the shower core (= point of the axis with the ground).
This means that $\tau_\mathrm{geo}$ can be determined experimentally by measuring the arrival time at each antenna, provided that the shower direction and the shower core are known accurately enough.

For a plane wavefront, $\tau_\mathrm{geo}$ is constantly 0. For any of the discussed wavefronts, which all are symmetric around the shower axis, $\tau_\mathrm{geo}$ obviously is a function of the 
distance $d$ to the shower axis.
Moreover, the antennas typically are located at a certain \lq{}height\rq{} $z_s$ in shower coordinates, where $z_s$ is the distance to the shower plane defined as the plane perpendicular to shower axis and containing the shower core. Generally $z_s$ is different from $0$. First, because the ground is not totally flat, second, because for inclined showers the shower plane is accordingly inclined against the ground. Thus, $\tau_\mathrm{geo}$ depends also on $z_s$.
Consequently, by definition $\tau_\mathrm{geo}$ is exactly 0 at the shower core ($d=0$ and $z_s=0$), but typically different elsewhere (see figure \ref{fig_hyperbola}).

For the hyperbolic wavefront we use the following parametrization describing the geometric delays as function of the coordinates ($d$ and $z_s$) and two parameters $\rho$ and $b$:
\begin{equation}
c \, \tau_\mathrm{geo}(d,z_s) = \\ \sqrt{(d\sin\rho)^2+(c\cdot b)^2} + z_s\cos\rho + c\cdot b
\label{eq_hyperbola}
\end{equation}
with the speed of light $c$. $\rho$ is the angle between the asymptotic cone of the hyperbola and a plane perpendicular to the shower axis, and $b$ is the offset of this cone to the hyperbola at the shower axis, i.e. at $d=0$.

Since the two-dimensional wavefront function $\tau_\mathrm{geo}(d,z_s)$ is difficult to illustrate, we also introduce a one-dimensional projection used for plotting (the numeric results in this paper are still obtained with the two-dimensional function).
For this purpose, the antenna positions are projected on the shower plane. The corresponding projected geometric delays are $c \, \tau_\mathrm{proj}(d) = c \, \tau_\mathrm{geo}(d,z_s) - z_s$.
For the parameter range studied at LOPES ($d < 200\,$m and $\theta < 45^\circ$) this is a very good approximation. Since $\rho$ is small, $z_s \cos \rho \approx z_s$, and the hyperbolic wavefront equation simplifies to:

\begin{equation}
c \, \tau_\mathrm{proj}(d) = \sqrt{(d\sin\rho)^2+(c\cdot b)^2} + c\cdot b
\label{eq_hyperbola_proj}
\end{equation}

For completeness, we also give the equations for the spherical and conical wavefront which have been used in earlier LOPES analyses.
Especially the conical wavefront might still be a useful simplification for larger antenna arrays, for which the wavefront shape within the first $50\,$m to the shower axis is irrelevant.

\begin{eqnarray}
c \, \tau_\mathrm{geo, sphere}(d,z_s) & = & \sqrt{(\frac{1}{\kappa}+z_s)^2 + d^2} - \frac{1}{\kappa} \\
                              & \approx & z_s + \frac{1}{2} \, \kappa \, (d^2 + \frac{z_s^2}{2}) \\
c \, \tau_\mathrm{proj, sphere}(d) & =  & \sqrt{1/\kappa^2 + d^2} -1/\kappa \\                             
c \, \tau_\mathrm{geo, cone}(d,z_s)   & = & d\sin\rho + z_s\cos\rho \\
                              & \approx & z_s + \rho \, d \\
c \, \tau_\mathrm{proj, cone}(d) & = & d\sin\rho
\end{eqnarray}
with the distance $d$ to the shower axis, the antenna height in shower coordinates $z_s$, the speed of light $c$, the curvature $\kappa$ (= reciprocal of the radius of curvature, respectively, the source distance $r_c$), and the cone angle $\rho$ between the cone and the shower plane.
The approximations apply for the conditions $\kappa d,\kappa z_s \ll 1$ and $\rho \ll 1$, respectively, which for LOPES is well fulfilled, since the typical distances $d$ and $z_s$ are at most a few $100\,$m, while the typical curvature radius $r_c=\kappa^{-1}$ is a few km, and the typical cone angle $\rho$ is below $2^\circ$.

In principle, also more detailed functions with additional parameters could be considered. E.g., in reference \cite{Lafebre2010}, a curved wavefront with four parameters is suggested. This might be mandatory, if the shape of the wavefront should be reproduced within a sub-nanosecond accuracy.
However, even testing a two-parameter function like the hyperbola is at the limit of what is possible with the precision of LOPES.
Furthermore, the simulations show that under ideal conditions the hyperbola is sufficient to achieve an $X_\mathrm{max}$ precision competitive to established techniques, and the practically achieved precision is limited by measurement uncertainties, but not by the wavefront function.
Consequently, at the moment there is no need to investigate wavefront functions more complex or detailed than the hyperbola.

\begin{figure}
  \centering
  \includegraphics[width=0.7\columnwidth]{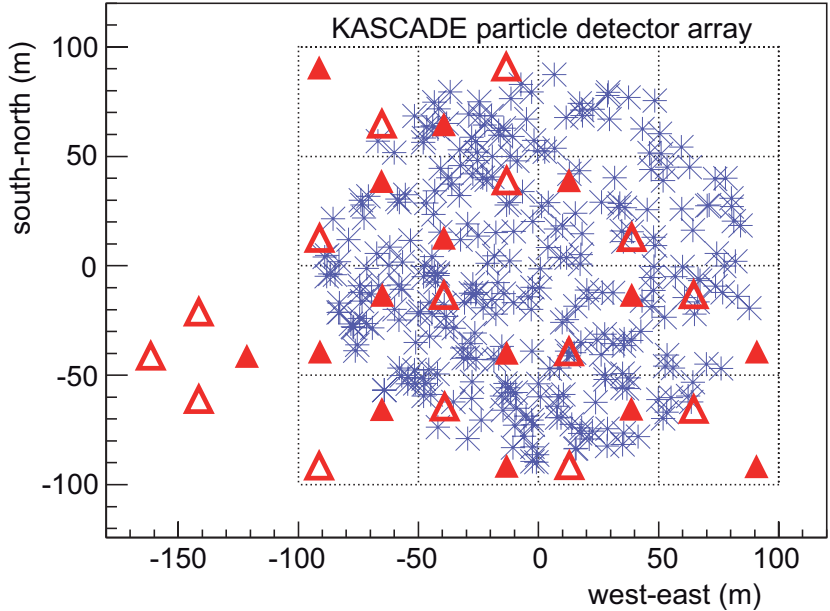}
  \caption{Map of the LOPES antennas aligned in east-west direction (triangles) and the shower cores of the measured events (stars); antennas indicated by open triangles have been switched off end of 2006 (to use them for polarization studies, instead). Thus, they are only taken into account for events measured in 2005 or 2006 (95 out of 316 events).}
   \label{fig_eventMap}
\end{figure}

\section{Data, Reconstruction, and Simulation}
The present analysis of the radio wavefront is based on 316 air shower events detected by both the KASCADE particle detector array \cite{AntoniApelBadea2003} and the LOPES digital radio interferometer \cite{FalckeNature2005}. Moreover, for each measured event we have performed two corresponding CoREAS \cite{HuegeCoREAS_ARENA2012} simulations, one for a proton as primary particle and one for an iron nucleus as primary particle. 

\subsection{Measurements}
The measurements have been performed with the LOPES antenna array, which digitally records the radio signal in the effective bandwidth between $43-74\,$MHz. A LOPES event is triggered and recorded whenever the KASCADE particle detector measured a high energy cosmic ray event. The detection threshold of LOPES depends on the arrival direction (e.g., the geomagnetic angle) and is in the order of $10^{17}\,$eV. Detailed descriptions of the LOPES experiment, its hardware and calibration procedures can be found in various references (see references \cite{HuegeARENA_LOPESSummary2010, SchroederLOPESsummaryARENA2012} for summaries of recent LOPES analyses). In the following sections, we will explain only the aspects of the LOPES experiment and the data reconstruction relevant for the wavefront analysis.

For this analysis, only measurements with east-west aligned LOPES antennas have been used, although LOPES contained also north-south and vertically aligned antennas for a part of the total operation time (2003-2013). The reason is that, due to the dominant geomagnetic origin of the radio emission, the signal-to-noise ratio is highest in the east-west aligned antennas for most shower geometries, which consequently provide the largest statistics. Moreover, we excluded data taken before 2005 because of the calibration quality and data after 2009, because then the antenna type was changed. 

The selection criteria for the used set of events are the same as in reference \cite{2013ApelLOPESlateralComparison}: in particular, the reconstructed energy has to be larger than $10^{17}\,$eV, the zenith angle smaller than $45^\circ$, the shower core has to lie within $90\,$m ground distance of the KASCADE center. 
Moreover, we applied several quality cuts on the radio signal to exclude background signals, e.g., we exclude events measured during thunderstorms, and require a minimum signal-to-noise ratio: the signal of all antennas is combined to a cross-correlation beam (cf. section \ref{sec_wavefrontReconstruction}). Its amplitude must be more than $14$ times larger than its standard deviation. For events with less than $30$ antennas contributing, the threshold is lowered proportional to the square root of the number of antennas. 
See figure \ref{fig_eventMap} for the core distribution of the remaining events. In contrast to the analysis described in reference \cite{2013ApelLOPESlateralComparison}, we did not use events which are reconstructed only by the high energy extension KASCADE-Grande, since the shower core of those events is not contained within the LOPES antenna array, which we considered mandatory for any wavefront reconstruction. This way, we obtained the sample of 316 measured LOPES events used for this analysis.

\subsection{CoREAS simulations}
The 316 measured events are accompanied by two sets of simulated events, one set with air showers initiated by protons, and one set with iron nuclei as primary particle. Each set is composed of one simulation for each measured event. As simulation program we have used CORSIKA \cite{HeckKnappCapdevielle1998} for the air showers, and the CoREAS extension \cite{HuegeCoREAS_ARENA2012} for the radio emission. CoREAS is based on the end-point formalism \cite{JamesEndPoint2010} and calculates the complete radio emission due to the acceleration, annihilation and production of charges. This means that CoREAS implicitly contains all emission mechanism which have been experimentally confirmed (cf.~section \ref{sec_introduction}).

The energy and geometry reconstructed by KASCADE for each event have been used as input for the simulations. Thus, the geometry and energy distribution of the measured and simulated events is the same, and the results can be compared directly. The calculation of the radio emission by the air shower is based on the particle cascade simulated by CORSIKA using the US standard atmosphere and the hadronic interaction models QGSJetII.03 \cite{OstapchenkoQGSjetII2006} and FLUKA \cite{BattistoniFLUKA2007}.

As the measurements, the simulations are affected by shower-to-shower fluctuations, which means that the shower maximum for each event will be different for the proton and iron simulations, and different to the unknown shower maximum of the measurements. Nevertheless, the statistics is large enough that the average characteristics are representative. In particular, the different simulations cover a range of $\rho$ and $X_\mathrm{max}$ which is wide enough to study correlations between both variables and with other shower parameters.

The CoREAS simulations are not affected by background and give us the exact arrival time at each antenna position within negligible numerical and sampling uncertainties. Furthermore, the arrival time at the shower core $t_0$ is extracted from the simulations. To make the simulations comparable to the measurements, we used only the east-west polarized part of the signal and applied a filter to the effective bandwidth of LOPES ($43-74\,$MHz). Consequently, the pulse arrival time used for the present analysis is the time when the east-west component of the filtered electric-field vector is maximum.

More details of the simulation procedure can be found in reference \cite{2013ApelLOPESlateralComparison}. There we used the same simulations to compare the lateral distribution of the radio amplitude to the measurements. The only discrepancy we found when comparing CoREAS to LOPES was the absolute scale of the amplitude: because of a still unknown reason, the measured amplitudes exceed simulations by a constant factor of approximately $2.5$ \cite{2013ApelLOPESlateralComparison}. All other tested features are compatible with the measurements, in particular the dependencies on the energy and shower geometry and the slope of the lateral distribution. For this reason, we have chosen CoREAS also for the present wavefront analysis. As a cross-check we repeated the analysis also with REAS 3.11 \cite{LudwigREAS3_2010} simulations, and obtained consistent results.

\begin{figure}
  \centering
  \includegraphics[width=0.7\columnwidth]{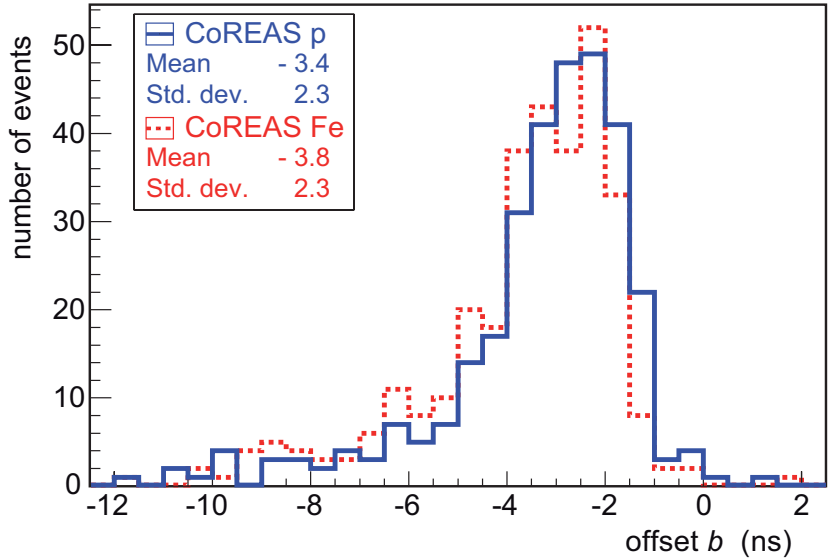}
  \caption{Distribution of the offset parameter $b$ determined by fitting a hyperbolic wavefront to the simulations. To decrease the number of free parameters in the wavefront fit, $b$ has been fixed to $-3\,$ns for the later analysis. This substantially improved the correlation of the fitted cone angle $\rho$ with the position of the shower maximum.}
   \label{fig_ConeOffset}
\end{figure}

\begin{figure*}
  \centering
  \includegraphics[width=0.49\columnwidth]{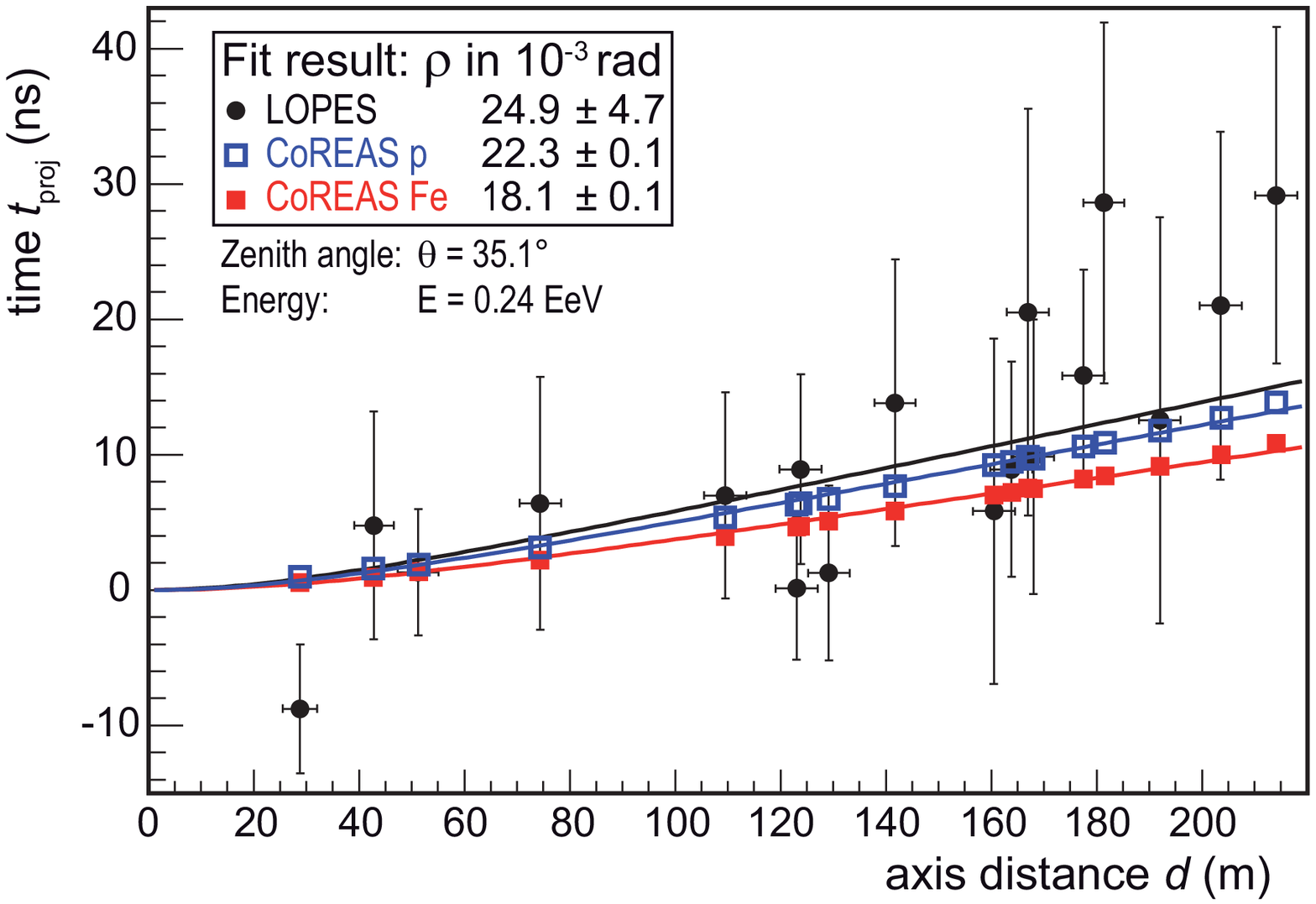}
  \includegraphics[width=0.49\columnwidth]{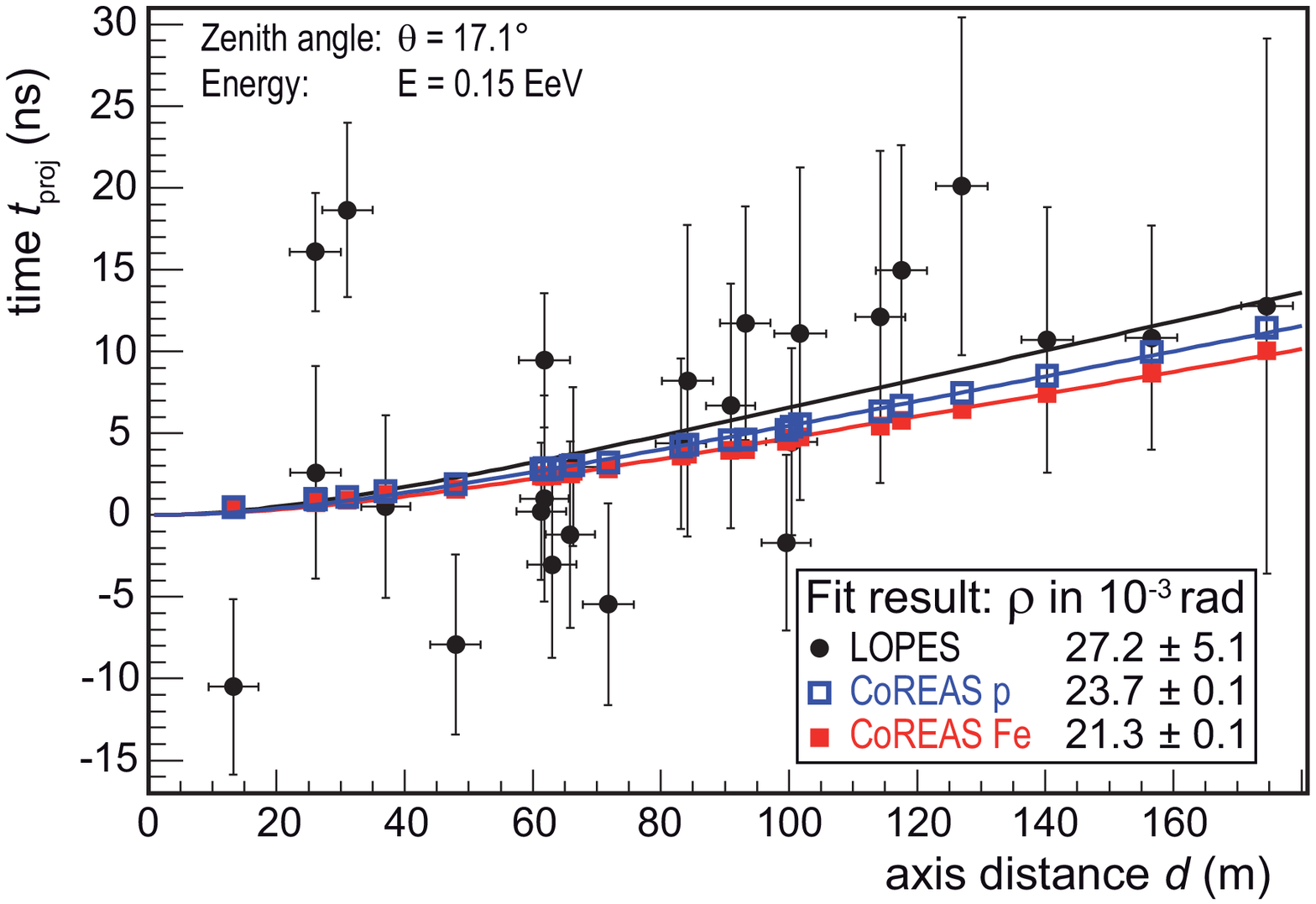}
  \caption{Two example events: arrival times and fitted hyperbolic wavefront (equation \ref{eq_hyperbola_proj} for $b$ fixed to $-3\,$ns) for the LOPES measurements and CoREAS simulations for protons and iron nuclei as primary particles; the measured and simulated arrival times (= time of the maximum) are corrected for the expected arrival time of an ideal plane wavefront, which would correspond to a horizontal line at $0\,$ns. Since the simulations are affected by shower-to-shower fluctuations, it is not expected that they would agree exactly with the measurements for individual events, but only on average. The x-error bars of the measurements are due to the uncertainty of the shower axis which is dominated by the core uncertainty; the y-error bars are dominated by the uncertainty due to noise, and include also the small calibration uncertainty.}
   \label{fig_example}
\end{figure*}

\subsection{Wavefront reconstruction}
\label{sec_wavefrontReconstruction}
For the reconstruction of the wavefront we used equation \ref{eq_hyperbola} for the LOPES measurements as well as for the simulations. However, the reconstruction process is different in the two cases. For the simulations, we fit the wavefront to the exactly known pulse arrival times at the individual antennas simulated for each event. For the measurements, we use an interferometric method, namely cross-correlation beamforming. The reasons for treating simulations and measurements differently are purely technical:

In contrast to the simulations, for the measurements $t_0$ is not known. Thus, it would have to be introduced as additional free parameter in a wavefront fit. This would considerably increase the fit uncertainties, in particular since $t_0$ would be correlated with $\rho$ in the fit. The beamforming technique only depends on the shape of the wavefront, but not on $t_0$. Consequently, it is ideal for the measurements. However, the beamforming technique would not bring any significant benefit for the simulations, since the wavefront can be determined directly by a fit. This provides the advantage that the statistical uncertainties of the wavefront parameters can be obtained from the fit, which unfortunately is not possible for the beamforming technique.

When fitting the hyperbolic wavefront to the simulations, in principle, we could perform the fit with two free parameters $b$ and $\rho$, but both would be correlated. Figure \ref{fig_ConeOffset} shows the distribution of $b$ for the simulations, when letting $b$ and $\rho$ free in the fit. However, when fixing $b$, the remaining free parameter $\rho$ has a higher sensitivity to the shower maximum, likely because the interdependences between $b$ and $\rho$ are suppressed in this case. Consequently, the wavefront is reconstructed with $\rho$ as only free fit parameter, and we fixed the offset parameter $b$ to its typical value of $-3\,$ns for both simulations and measurements. For fixing $b$, the exact value is less important: fixing $b$ to $-2\,$ns or $-4\,$ns, instead, the finally obtained $X_\mathrm{max}$ resolution changes by less than $15\,\%$, and is still almost twice as good than for a freely fitted $b$.

Figure \ref{fig_example} shows the reconstructed wavefront for two example events. For illustration also the experimental data are shown in this example. However, we did not use the shown fit to reconstruct $\rho$, but the beamforming technique for the measurements. This yields very similar results, though, but is on average more accurate.

The cross-correlation beamforming procedure used in LOPES is described in several references \cite{FalckeNature2005, HuegeARENA_LOPESSummary2010}. The digital recorded traces, i.e.~the measured field strength as function of time, are digitally shifted in time according to the geometric delays, in this case calculated with equation \ref{eq_hyperbola}. For this purpose, the shower core is set to the value of the KASCADE reconstruction. The shower direction, however, is implicitly determined by the beamforming, since the antenna coordinates $d$ and $z_s$ depend on the shower axis. In a first step, we evaluate a three dimensional grid $(d, z_s, \rho)$, and in a second step, we maximize the cross-correlation amplitude by optimizing the free parameters with a simplex fit. 
To speed up computing time, we use the KASCADE values for the shower axis as starting point, but we have checked that our grid is large and dense enough to avoid a systematic bias. The exact algorithms for the reconstruction process can be obtained directly from our reconstruction software, which is available as open source \cite{CRtools}.

\subsection{Uncertainties}
While for the simulations, most uncertainties are negligible, they play a significant role for the LOPES measurements. There are three relevant types of uncertainties for the arrival time measurements. First, uncertainties on the determined arrival times themselves, in particular due to background. Second, uncertainties on the time when the wavefront hits the ground at the impact point of the shower axis, i.e. uncertainties on $t_0$. Third, uncertainties on the distance of each individual antenna to the shower axis. Finally, there is a systematic uncertainty due to the asymmetry of the wavefront, but the effect is negligible against the measurement uncertainties of LOPES.

\subsubsection{Arrival time measurement}
Since LOPES features a ns-precise relative time calibration \cite{SchroederTimeCalibration2010}, calibration uncertainties are negligible for typical measurements. Also possible uncertainties due to the sampling are negligible, since the sampling frequency of LOPES ($80\,$MHz) is more than twice the effective bandwidth ($43-74\,$MHz). Hence, according to the Nyquist theorem, the measured radio signal can be fully reconstructed even in between the measured samples. 
For the present LOPES analysis, this is done using zero-padding up-sampling by a factor of 16, i.e., to sub-ns precision. We experimentally tested the quality of this up-sampling procedure, by feeding defined calibration pulses into our data acquisition at different times relative to the original sampling. 
After up-sampling, we indeed obtained the same pulse shape to better than ns-precision, independent of how the pulse fell relative to the ADC sampling. Consequently, for a typical measurement, neither the time calibration nor the sampling frequency of LOPES contribute significantly to the arrival time uncertainty.

Instead, for typical signal-to-noise ratios the major source of uncertainty is noise, and arrival time uncertainties due to noise typically reach a few ns. We have determined these uncertainties for the pulse time in individual antennas with a small Monte Carlo simulation: By adding real, measured noise to measured pulses from a pulse generator, we could study with high statistics how the measured pulse time depends on the signal-to-noise ratio. Details can be found in reference \cite{SchroederNoise2010}. The conclusion is that for air shower measurements with LOPES, the arrival time uncertainty is between $2\,$ and $17\,$ns, depending on the signal-to-noise ratio in the individual measurement.

\subsubsection{$t_0$ of the wavefront}
Theoretically, $t_0$ is well defined by the time when the radio wavefront touches ground at the shower core (= intersection point of the shower axis with the ground plane). For the simulations, $t_0$ can be obtained directly by simulating an antenna at the shower core (only for technical reasons we simulated antennas at $1\,$cm distance to the core). For the measurements, $t_0$ cannot be determined directly, since the probability that the shower axis hits directly one antenna is too low. 
In addition, the position of the shower core is not known exactly. Moreover, the arrival time measurement at any antenna, including those close to the shower axis, come with a significant uncertainty due to noise. For these reasons, a measurement of $t_0$ is challenging and might only be achieved with extremely dense antenna arrays like LOFAR \cite{SchellartLOFAR2013}. 
The difficulty to obtain $t_0$ for the measurements is the principle reason, why we did not fit the wavefront directly to the measured arrival time distribution, but instead determined it by beamforming, because for beamforming only the shape of the wavefront is relevant, but not the absolute pulse time.

\subsubsection{Geometry}
Since the wavefront depends on the arrival time as a function of distance to the shower axis, not only time uncertainties matter, but also geometry uncertainties. However, at LOPES the geometric uncertainties are relatively small compared to the large uncertainties due to noise. The antenna positions relative to each other have been measured with differential GPS with a precision of approximately $5\,$cm, corresponding to less than $0.2\,$ns propagation time for the radio signal. 
The uncertainty of the shower core is reconstructed by the dense KASCADE particle detector array with an accuracy of better than $4\,$m \cite{AntoniApelBadea2003}. This corresponds to correlated errors in the arrival time of about $7\,$ns for the average zenith angle of approximately $30^\circ$.
The relative influence of the geometry uncertainty in comparison to the arrival time uncertainty can be judged from figure \ref{fig_example} in which the geometry uncertainties are indicated as x-error bars, and the arrival time uncertainties are indicated as y-error bars. Thus, the geometry uncertainty generally is smaller than the time uncertainty due to noise, but not completely negligible. It might become more important for arrays with lower background level and larger antenna spacing.

\begin{figure}
  \centering
  \includegraphics[width=0.7\columnwidth]{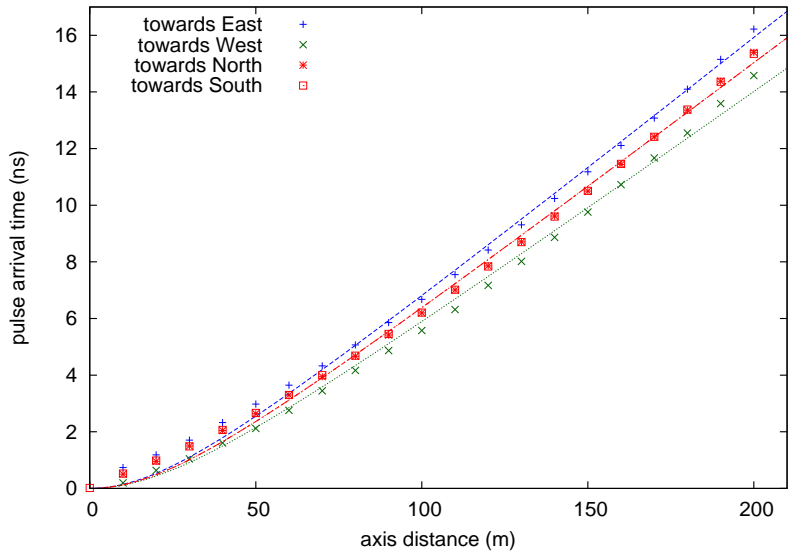}
  \caption{Arrival times of the pulse maximum of a typical CoREAS simulation for a vertical shower initiated by a proton with $10^{17}\,$eV, and fits of a hyperbolic wavefront, separately for antennas in the east, west, north and south direction. The wavefront shows an east-west asymmetry, and it slightly deviates from an ideal hyperbolic shape. However, the size of both effects is small and irrelevant compared to the measurement uncertainties of LOPES of several nanoseconds.}
   \label{fig_wavefrontAssymetry}
\end{figure}

\subsubsection{Asymmetry of the wavefront}
\label{sec_assymetry}
The amplitude of the radio signal is not totally symmetric around the shower axis, as has been shown experimentally \cite{CODALEMAMarinICRC2011, NellesLOFAR_LDF2014}: For the simple case of a vertical shower, the geomagnetic contribution is purely east-west polarized and interferes with the radial polarized contribution of the Askaryan effect. Until now, the effect on the arrival times, i.e., the asymmetry of the wavefront, has not been studied. Since it turns out that the effect is smaller than the measurement accuracy of LOPES, we analyzed the asymmetry with simulations made for a simple geometry. 

We used very detailed CoREAS simulations of vertical showers initiated by primary protons and iron nuclei with an energy of $10^{17}\,$eV made for the geomagnetic field and the altitude of LOPES. Instead of the LOPES antenna geometry, we simulated an artificial antenna grid placing antennas exactly on the north-south and east-west axis with $10\,$m spacing up to a distance of $200\,$m. As for the other simulations used in this paper, we determined the arrival time in each antenna as the time at which the field strength of the east-west polarization is maximum, after applying a filter to the effective bandwidth of LOPES. Then, we fitted the hyperbolic wavefront to the arrival times using a fixed offset parameter $b = -3\,$ns, and compared the result for $\rho$ in east, west, north and south direction (see figure \ref{fig_wavefrontAssymetry}). The figure also illustrates that the hyperbolic wavefront only approximates the individual arrival times, but the deviations are on a sub-nanosecond level, i.e., smaller 
than the achieved timing precision.

As expected, the arrival times and the fitted wavefront are approximately equal in north and south direction, i.e., for all simulated events the values for $\rho$ are equal to better than $0.5\,\%$. However, in the case of constructive or destructive interference, i.e., in east and west direction, there is a significant difference.  Towards East the radio signal arrives slightly later, and towards West slightly earlier, at least when determining the arrival time by the pulse maximum as explained above. The maximum size of the effect is in the order of the time calibration uncertainty, i.e., $1\,$ns, and therefore is smaller than the typical timing uncertainty due to noise, which amounts to a few ns.

The effect on $\rho$ is in the order of $\pm 5\,\%$, i.e., the value of $\rho$ in east direction is approximately $5\,\%$ larger than in north or south direction, and approximately $10\,\%$ larger than in west direction. However, in a realistic scenario there will always be antennas in different directions from the shower axis and the effect of the asymmetry will partly average out. Thus, the net effect in a real measured event will typically be much smaller. Still, the asymmetry will introduce some scatter of the individual arrival times around the idealized symmetric hyperboloid, which contributes to the fit uncertainty of $\rho$. For the full data set of the CoREAS simulations made for the realistic sample of shower geometries and the LOPES antenna positions, the average uncertainty of $\rho$ is $4.6\,\%$. This uncertainty includes the systematic uncertainty due to the asymmetry which, consequently, might be a significant fraction of the $4.6\,\%$, but not more than this.

Concluding, the asymmetry of the wavefront might be important when aiming at a measurement precision of better than $5\,\%$ for the cone angle $\rho$, which should result in a precision for $X_\mathrm{max}$ better than the $25\,$g/cm\textsuperscript{2} which we achieve in the simulations without correcting for the asymmetry. With respect to the experimental uncertainties of LOPES, the effect of the asymmetry is negligible. Hence, it is ignored for the following analysis.

\begin{figure}
  \centering
  \includegraphics[width=0.7\columnwidth]{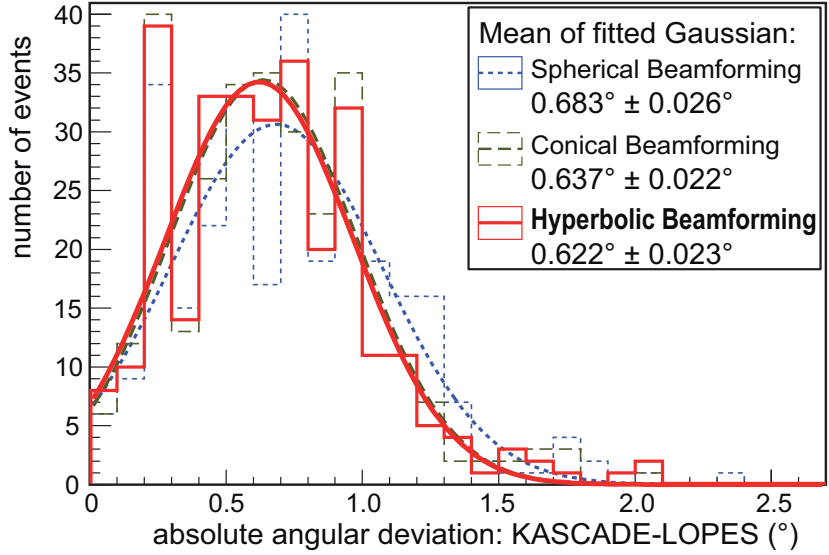}
  \caption{Deviation between the arrival direction reconstructed with the KASCADE particle detector array and reconstructed with LOPES using three different wavefront shapes for the beamforming analysis (corresponding to the three shown histograms and fitted Gaussians). The mean deviation can be taken as upper limit for the direction precision of LOPES for air showers, which consequently is best when using a hyperbolic wavefront for the cross-correlation beamforming.}
   \label{fig_angularResolution}
\end{figure}

\begin{figure}
  \centering
  \includegraphics[width=0.7\columnwidth]{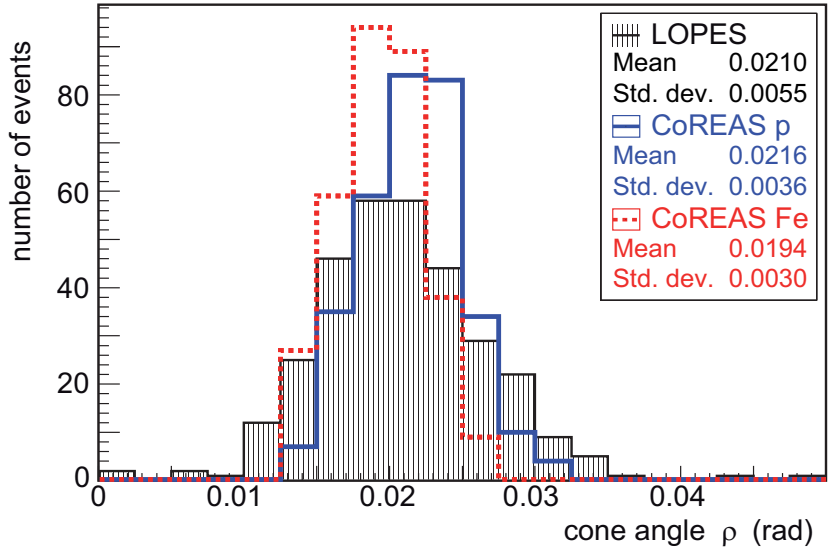}
  \caption{Distribution of the cone angle $\rho$, obtained by hyperbolic beamforming for the measured events, and by fitting the wavefront function for the simulated events. The range of the measured cone angles corresponds to approximately $0.6^\circ - 2.0^\circ$.}
   \label{fig_coneAngleComparison}
\end{figure}

\begin{table}
\centering
\caption{Mean and standard deviation of the reduced $\chi^2$ obtained with the different wavefront shapes for the CoREAS simulations with iron and proton primaries. For the LOPES measurements, no comparable $\chi^2$ can be given, since the wavefront is determined by beamforming and not by a fit.}
\label{tab_redChi2comparison}
\vspace{0.1 cm}
\begin{tabular}{lcc}
wavefront shape & \multicolumn{2}{c}{reduced $\chi^2$ for}\\
used for fitting & proton sims. & iron sims.\\
\hline
Sphere                   &~~$0.82 \pm 0.67$~~&~~$0.52 \pm 0.48$~~\\
Cone                     &~~$0.37 \pm 0.23$~~&~~$0.33 \pm 0.17$~~\\
Hyperbola ($b=-3\,$ns)~~~&~~$0.07 \pm 0.07$~~&~~$0.04 \pm 0.04$~~\\
\hline
\end{tabular}
\end{table}

\begin{figure}
  \centering
  \includegraphics[width=0.6\columnwidth]{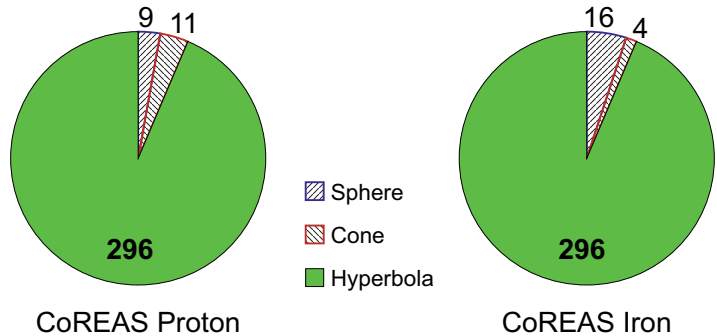}
  \caption{Number of simulated events for which the fit of a certain wavefront shape gives the lowest $\chi^2$. For the hyperbolic fit, the offset parameter has been fixed to $b = -3\,$ns, such that all three wavefront fits have the same number of degrees of freedom.}
   \label{fig_compareWavefrontsChi2}
\end{figure}

\section{Results}

\subsection{Wavefront shape}
We found several indications that a hyperbolic wavefront provides a better description than a spherical or conical wavefront. A simple plane wavefront has already been excluded in an earlier analysis \cite{NiglDirection2008}. While for the simulations the situation is clear, the measurements are affected by the large background level and we can only find hints, but no definite proof that a hyperbola comes closest to reality. 

The experimental indications for the real wavefront shape are of indirect nature. The arrival direction reconstructed with LOPES is slightly closer to the direction reconstructed with KASCADE when using a hyperbolic wavefront instead of a spherical or conical wavefront (figure \ref{fig_angularResolution}). Although the improvement in the direction precision is small, it is significant: For $(72 \pm 5)\,\%$ of the events, the hyperbolic beamforming comes closer to the KASCADE direction than the spherical beamforming, and for $(58 \pm 4)\,\%$ of the events the hyperbolic beamforming is better than the conical one. Furthermore, the signal-to-noise ratio of the cross-correlation amplitude is best for hyperbolic beamforming. But again the difference is small, and of statistical significance only in the comparison to conical beamforming: for hyperbolic beamforming, the cross-correlation amplitude has a better signal-to-noise ratio compared to spherical beamforming for $(52 \pm 4)\,\%$ of the events, and compared 
to conical beamforming for $(60 \pm 4)\,\%$ of the events.

This means that the hyperbolic wavefront is favored over both, the spherical and the conical wavefront. However, when comparing the spherical and the conical wavefront with each other, the result of the comparison is not so clear. With respect to the arrival direction, the conical wavefront is significantly favored over the spherical one, but when comparing the signal-to-noise ratio, the spherical wavefront is slightly favored over the conical one.

For the simulations, the comparison is much easier, since the quality of the different wavefront fits can be compared directly, e.g., by comparing the reduced $\chi^2$. The absolute value of the reduced $\chi^2$ is not meaningful, since the individual arrival times have no statistical uncertainty for the simulation. Still, the $\chi^2$ values can be used to compare the different fits with each other. In particular, since $b$ has been fixed to $-3\,$ns for the hyperbolic wavefront, the number of free parameters is equal for all three cases, and the $\chi^2$ for all three fits (spherical, conical and hyperbolic wavefront) can be compared directly. The mean and standard deviation of the reduced $\chi^2$ for the different wavefronts are stated in table \ref{tab_redChi2comparison}. Moreover, an event-by-event comparison shows that the hyperbolic wavefront gives the best fit for approximately $94\,\%$ of the events (figure \ref{fig_compareWavefrontsChi2}), even though the offset parameter $b$ of the hyperbola has 
been fixed, and thus is not optimal for each individual event.


Finally, we found no indication that the simulations would contradict the measurements. In particular, the distribution of the cone angle $\rho$ is compatible (figure \ref{fig_coneAngleComparison}). Considering these different indications, we decided to perform the remaining analyses only with the cone angle $\rho$ determined from a hyperbolic wavefront. Nevertheless, a conical or spherical wavefront might be a sufficient approximation for some applications.

\begin{figure}
  \centering
  \includegraphics[width=0.7\columnwidth]{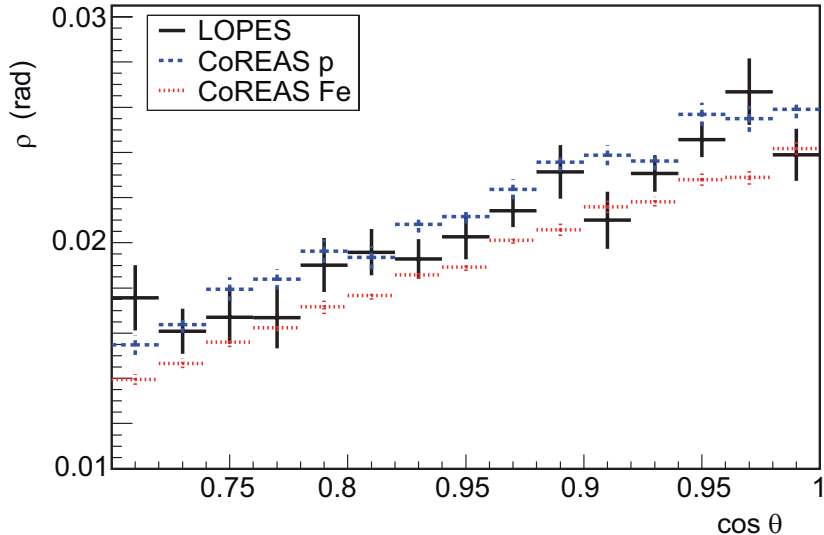}
  \caption{Dependence of the cone angle $\rho$ on the zenith angle $\theta$: mean values of $\rho$ and statistical uncertainties for equidistant bins of $\cos \theta$. For both, measurements and simulations, inclined showers show on average a flatter wavefront. This can be explained by the larger distance of the detector to the shower maximum.}
   \label{fig_zenithDependence}
\end{figure}

\begin{figure}
  \centering
  \includegraphics[width=0.7\columnwidth]{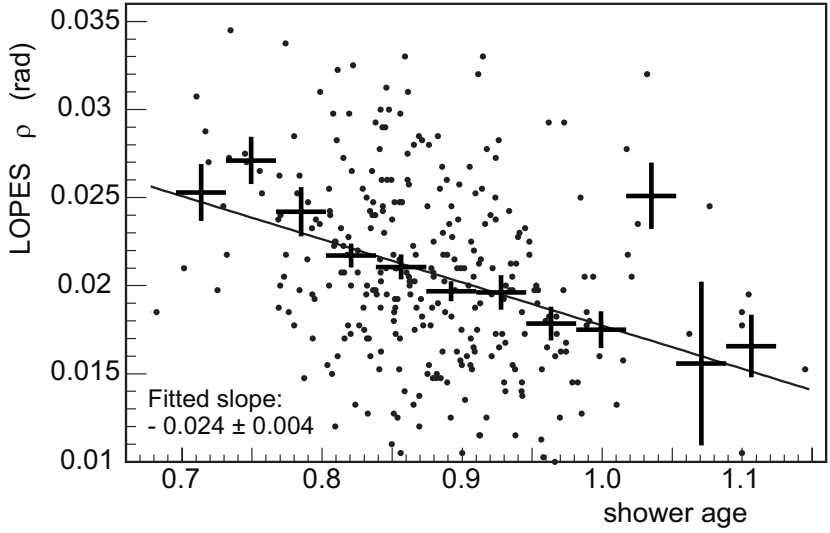}
  \caption{Correlation of the cone angle $\rho$ measured with LOPES and the shower age parameter reconstructed by KASCADE; individual data points and a profile (horizontal bars are at the position of the mean in each bin, where the bin width is the size of the horizontal bars; the size of the vertical bars is the standard deviation divided by the square root of the number of points in each bin). Older showers with a large age typically have a more distant shower maximum, and thus are expected to show a flatter wavefront, i.e., a smaller cone angle $\rho$. The correlation is weak, but the slope of a line fitted to the individual data points is significantly different from 0. Unfortunately, the beamforming method does not provide uncertainties for the measured data points. Thus, equal weight is given to all points for the fit.}
   \label{fig_ageCorrelation}
\end{figure}

\begin{figure*}
  \centering
  \includegraphics[width=0.49\columnwidth]{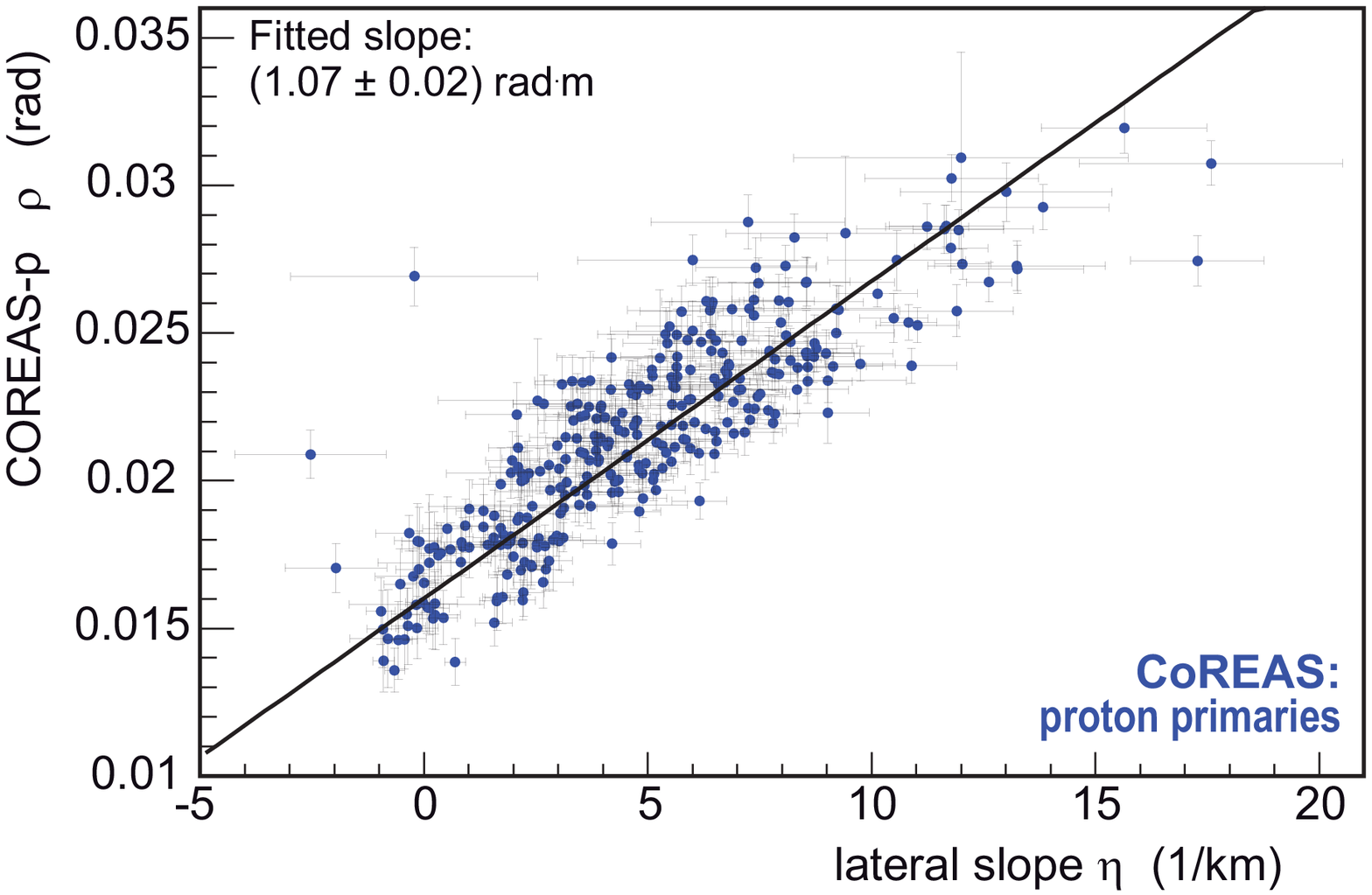}
  \includegraphics[width=0.49\columnwidth]{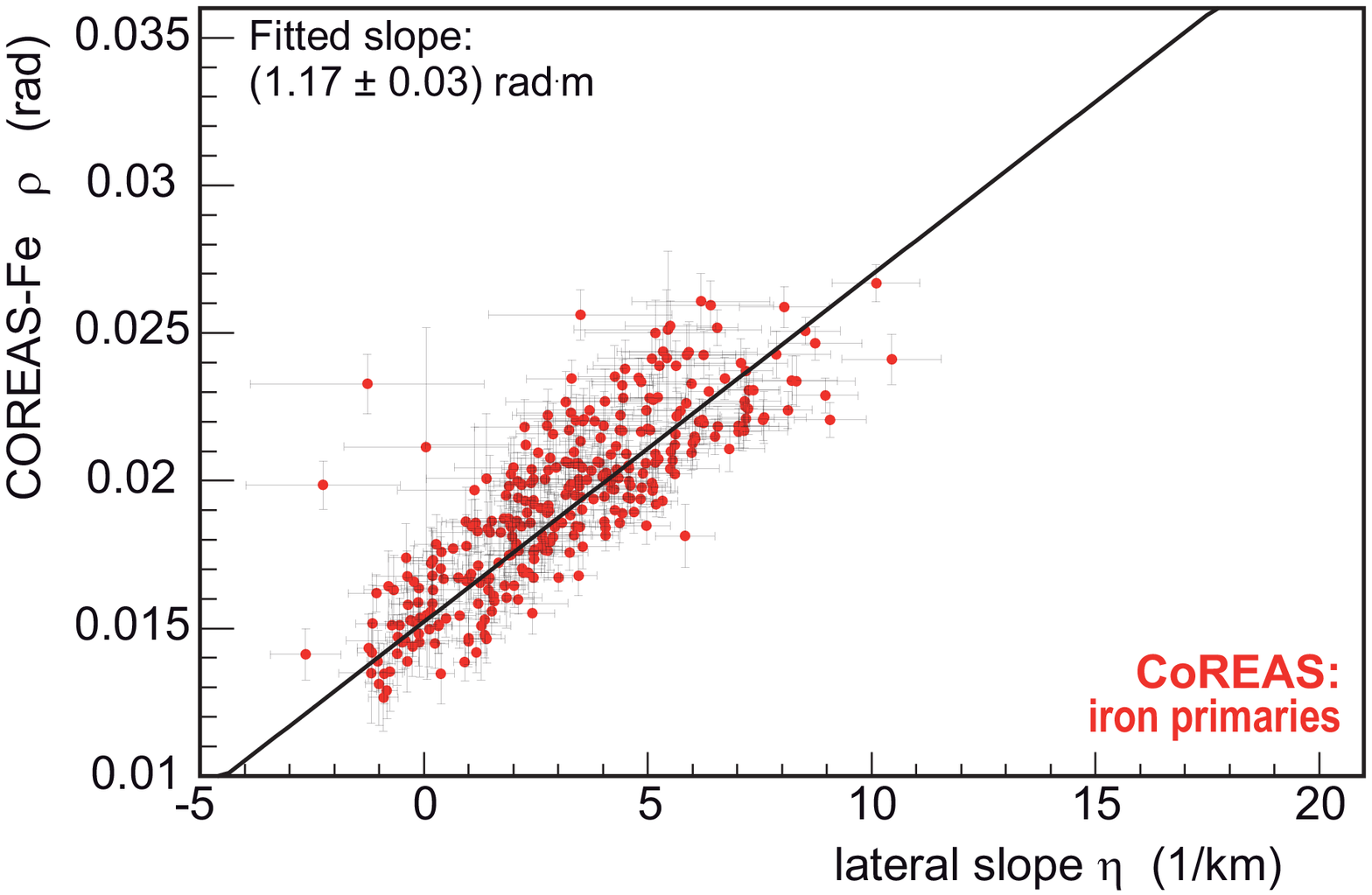}
  \caption{Correlation of the cone angle $\rho$ and the slope parameter $\eta$ of the lateral distribution for the CoREAS proton simulations (left) and the iron simulations (right). As expected, showers with a steeper lateral distribution (large $\eta$) also have a steeper radio wavefront (large $\rho$), since both are affected by the distance to the shower maximum.}
   \label{fig_RhoVsEtaCoreas}
\end{figure*}

\begin{figure}
  \centering
  \includegraphics[width=0.7\columnwidth]{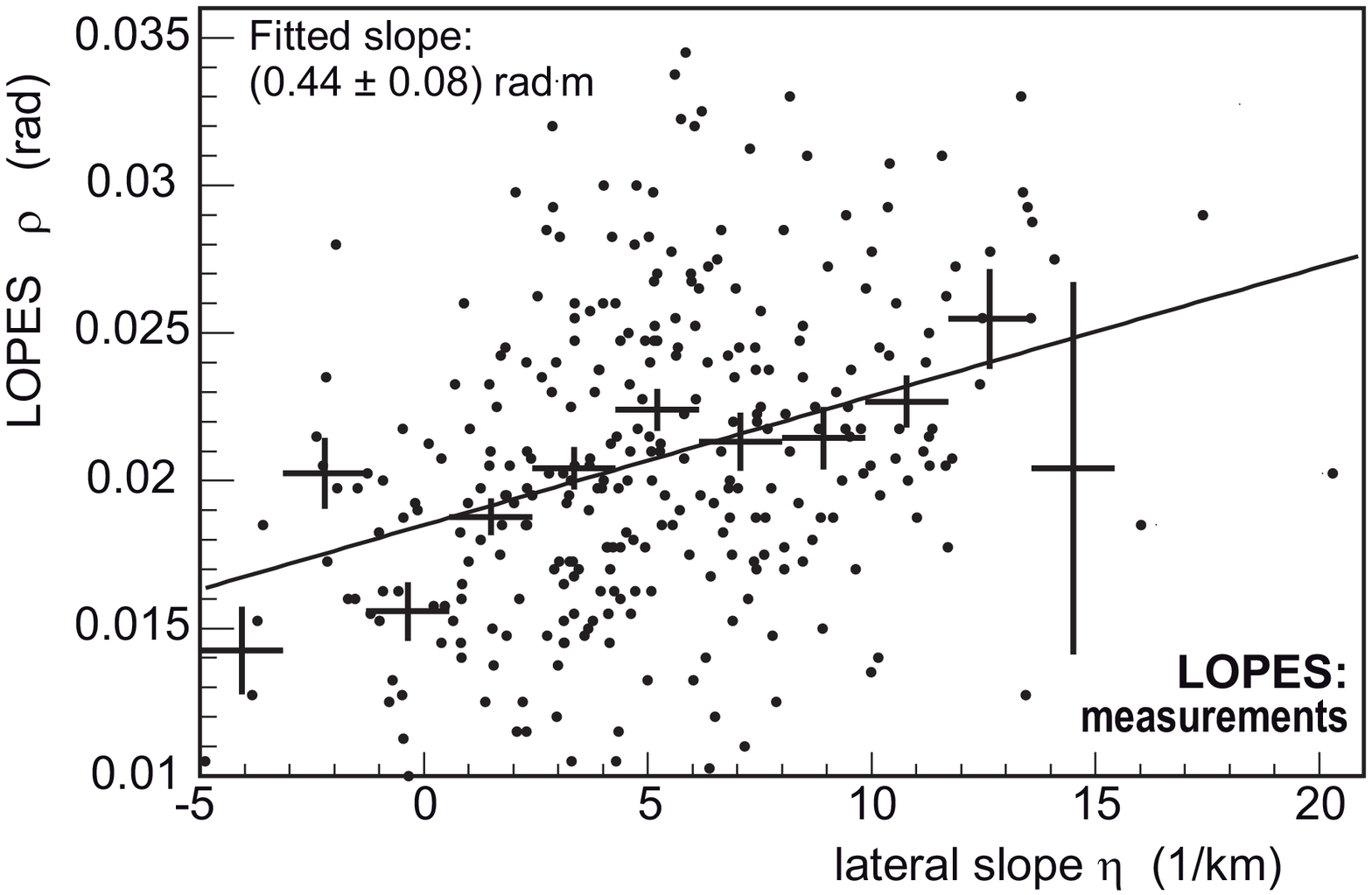}
  \caption{Correlation between the cone angle $\rho$ and the slope parameter $\eta$ of the lateral distribution for the LOPES measurements; individual data points and a profile. The measured correlation shows the same trend as the simulations (figure \ref{fig_RhoVsEtaCoreas}), but the line fitted to the individual data points indicates a different slope. Probably, this is caused by the large measurement uncertainties causing a bias on the slope of the correlation. Unfortunately, the beamforming method does not provide uncertainties for the measured data points. Thus, equal weight is given to all points for the fit.}
   \label{fig_RhoVsEtaLOPES}
\end{figure}

\subsection{Correlation with shower parameters}
We have studied various correlations of the cone angle $\rho$ with different parameters of the same air showers. We find several indications that the steepness of the radio wavefront depends in first oder on the geometric distance of the shower maximum to the detector and consequently, on the shower development. However, the simplest geometric approximation of a radio point source in the shower maximum fails, since the wavefront is not spherical, but hyperbolic. Thus, a deep quantitative understanding of the correlations between the wavefront shape and the shower parameters is non-trivial and will depend on the details when modeling the radio emission. Unfortunately, LOPES does not feature the required measurement precision for detailed tests. Still, we see the expected correlations in our data, and will discuss them on a qualitative level, which is sufficient for our phenomenological approach for the reconstruction of the shower maximum with the radio wavefront: the more distant the shower maximum from the 
detector, the flatter the radio wavefront, i.e.~the smaller the cone angle.

\begin{figure*}
  \centering
  \includegraphics[width=0.49\columnwidth]{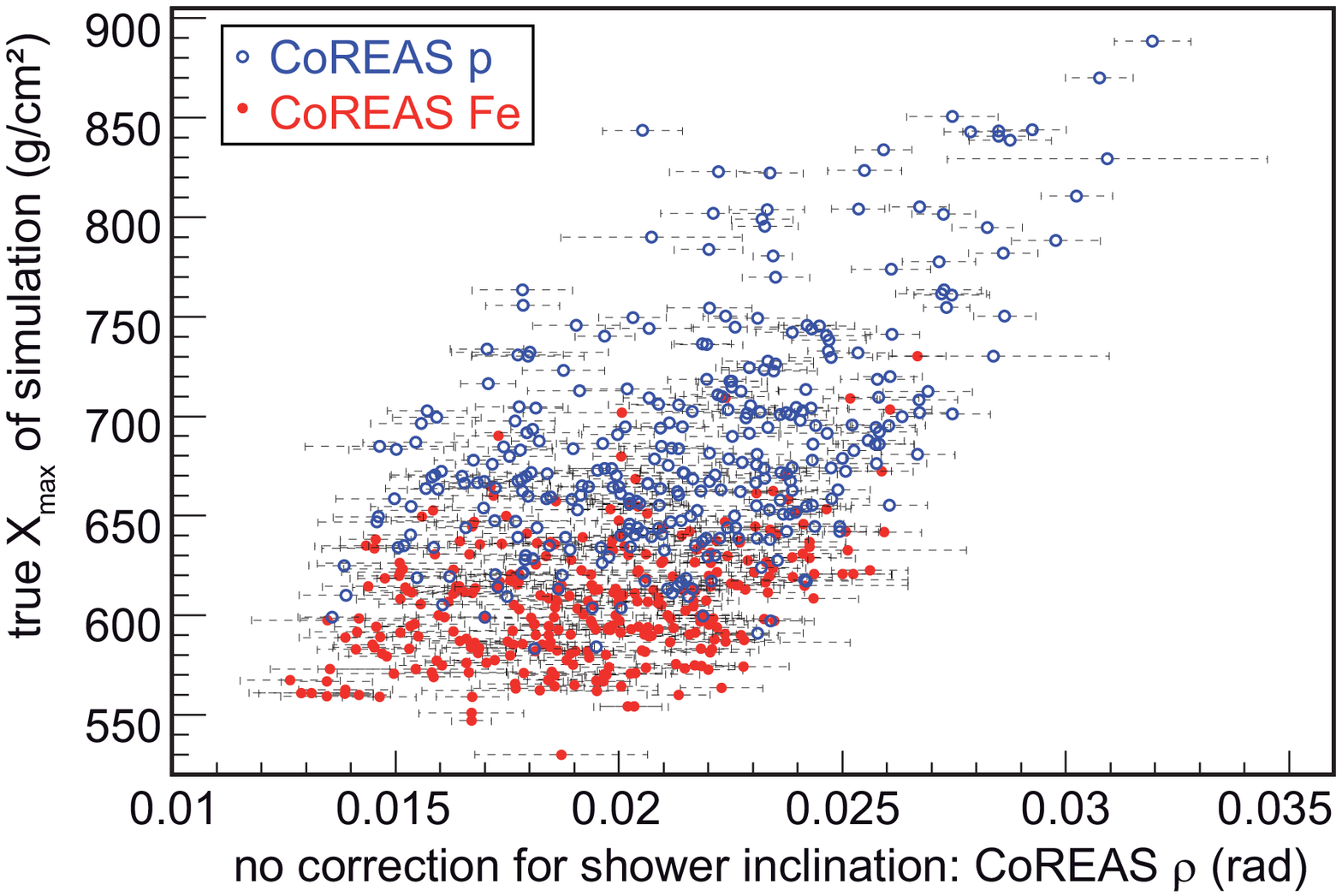}
  \includegraphics[width=0.49\columnwidth]{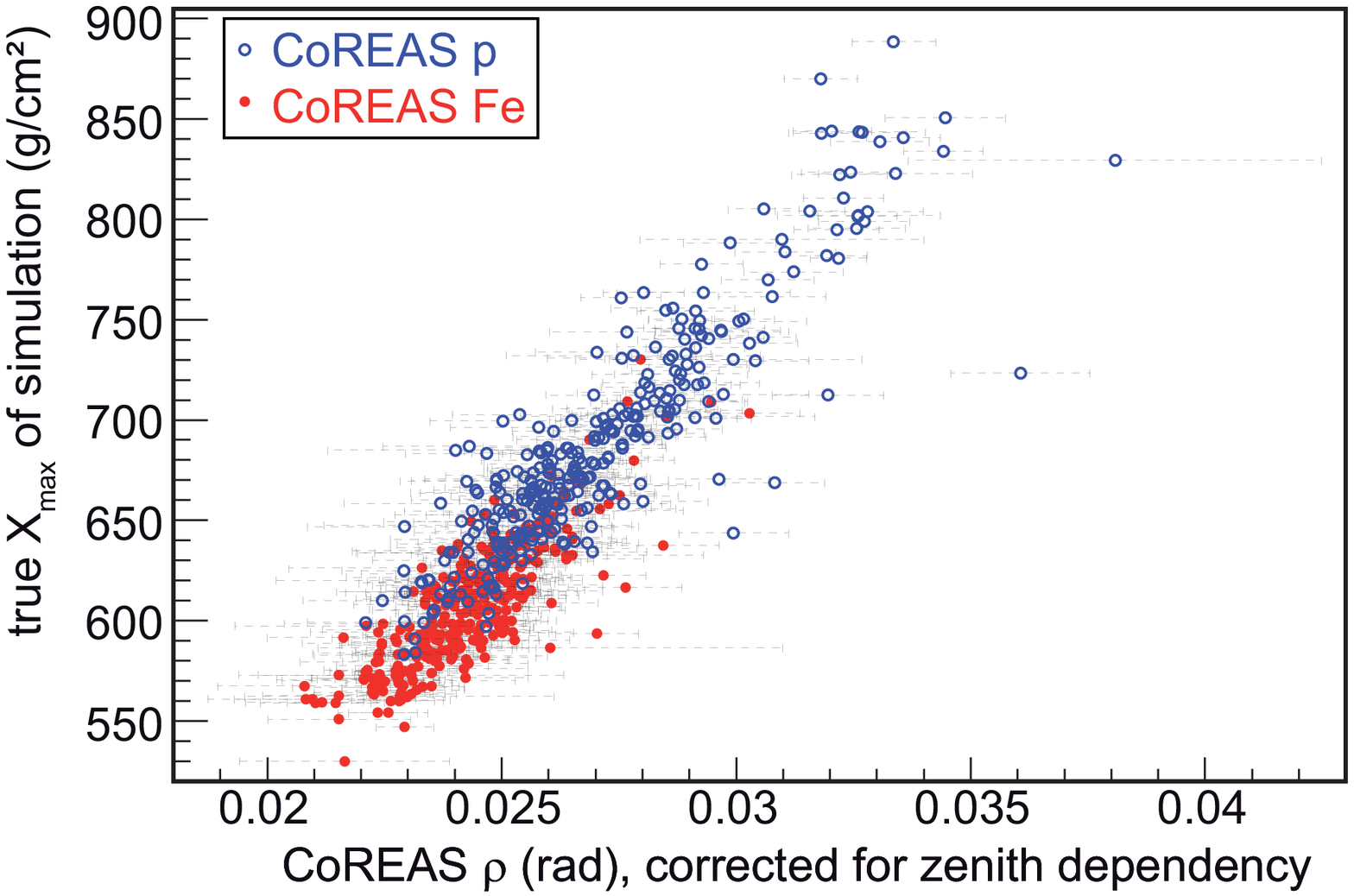}
  \caption{Correlation of the cone angle $\rho$ with the atmospheric depth of the shower maximum $X_\mathrm{max}$ before (left) and after (right) correction for the zenith angle; the systematic difference between the proton and iron simulations is small compared to the average spread (cf.~figure \ref{fig_XmaxDeviation}).}
   \label{fig_trueXmaxCorrelation}
\end{figure*}

Consequently, the cone angle is correlated with the shower inclination (figure \ref{fig_zenithDependence}), since the distance of the shower maximum to the detector depends not only on the atmospheric depth of the shower maximum, $X_\mathrm{max}$, but also on the zenith angle $\theta$. Moreover, the measured cone angle $\rho$ is correlated with the shower age reconstructed from the KASCADE array (figure \ref{fig_ageCorrelation}). Due to the large measurement uncertainties, the correlation is weak, but still statistically significant, since the slope of a fitted line is inconsistent with 0. The value of the fitted slope depends slightly on the fit method, e.g., if the profile or the individual points are fitted. However, the qualitative result, namely that there is a significant correlation, is robust. For the simulations, the KASCADE age parameter is not available, but the true position of the shower maximum is, which is an even better observable to test the correlation of $\rho$ with the longitudinal shower 
development (see figure \ref{fig_trueXmaxCorrelation} in the next section).

Finally, we expect a correlation of the radio wavefront with another radio observable, namely the slope of the lateral distribution, which is sensitive to the shower development, too \cite{2012ApelLOPES_MTD}. In reference \cite{2014ApelLOPES_MassComposition}, we have shown that the lateral distribution of the radio emission can better be described by a Gaussian function with three parameter than with the previously used exponential function: $\epsilon \propto \exp(- \eta \cdot d)$, with the amplitude $\epsilon$, the axis distance $d$ and the slope parameter $\eta$. Still, due to the large measurement uncertainties, the exponential function gives a sufficient fit to the majority of LOPES events, and offers the advantage that its two parameters can be understood more intuitively: one parameter, namely the proportionality constant, determines the amplitude scale and is almost linearly correlated with the primary energy \cite{2010ApelLOPESlateral}, the other parameter, $\eta$, determines the slope, and is 
correlated with the distance to the shower maximum \cite{2012ApelLOPES_MTD}:

The smaller this slope parameter $\eta$, the flatter the lateral distribution, and the more distant the shower maximum. Indeed, we see a correlation between $\eta$ and $\rho$ for both simulations (figure \ref{fig_RhoVsEtaCoreas}) and measurements (figure \ref{fig_RhoVsEtaLOPES}). The parameter range and the trend is the same for both, but a fitted line shows a steeper slope for the simulations. The value for the fitted slope and the difference between simulations and measurements vary only slightly (on a level of $10\,\%-20\,\%$) with the chosen fit method, i.e., if individual error bars are considered as weights, or if the fit is performed on the individual points or on the profile.

The apparent deviation between the measurements and simulations can be explained by the large uncertainties in the measurement, since the additional spread due to the uncertainties causes a systematic bias on the slope of the fitted line. Although the individual measurement uncertainties cannot be determined for the cross-correlation beamforming, the spread gives an impression of the size of the uncertainties. When artificially increasing the 
spread of the simulations by adding an additional Gaussian uncertainty of the expected order of magnitude, the slope of the correlation indeed changes significantly. Consequently, the different slope does not constitute an incompatibility between measurements and simulations. Instead, it illustrates the significant uncertainties for LOPES, and is the principal reason why we can confirm the predicted and expected correlation only on a qualitative level.

\subsection{Reconstruction of the shower maximum}
Since the wavefront shape, namely the cone angle $\rho$, depends on the longitudinal shower development, it can be used to reconstruct the atmospheric depth of the shower maximum $X_\mathrm{max}$. For this purpose, the zenith dependence (figure \ref{fig_zenithDependence}) has to be corrected to extract the remaining sensitivity on $X_\mathrm{max}$.

We have chosen a phenomenological approach for this correction. In principle, also an analytical approach of the geometric dependence might be feasible. However, an analytic approach would be highly non-trivial, because the fact that the wavefront is not spherical implies that it is insufficient to assume a point source at the shower maximum for the geometry correction. Thus, we have determined the dependence of $\rho$ on $\cos \theta$ by a power law fit to the dependence observed in the simulations (cf.~figure \ref{fig_zenithDependence}), and use the following equation for the correction of this zenith dependency:

\begin{equation}
\rho_\mathrm{cor} = \rho \cdot \cos^{-\gamma} \theta
\label{eq_zenithcorrection}
\end{equation}

with the zenith angle $\theta$ and a power law index $\gamma$ determined by a fit. We found $\gamma = 1.43 \pm 0.02$ for the proton simulations, and $\gamma = 1.55 \pm 0.03$ for the iron simulations. For the LOPES measurements, $\gamma$ cannot be determined by a fit to the individual events, since the uncertainties are too large. A fit to the profile shown in figure \ref{fig_zenithDependence} yields $\gamma = 1.42 \pm 0.14$. Thus, the experimentally observed zenith dependence of $\rho$ is in accordance with the simulations. For this reason, we have decided to use the average $\gamma$ of the proton and iron simulations, namely $\gamma = 1.485$, to correct all simulations and the measurements.

Figure \ref{fig_trueXmaxCorrelation} shows that after this correction of the zenith dependence, there remains a strong and approximately linear correlation of $\rho_\mathrm{cor}$ with $X_\mathrm{max}$. Consequently, $X_\mathrm{max}$ can be determined with a simple linear equation:

\begin{equation}
X_\mathrm{max} = const \cdot \rho_\mathrm{cor}
\label{eq_XmaxReconstruction}
\end{equation}

\begin{figure}
  \centering
  \includegraphics[width=0.7\columnwidth]{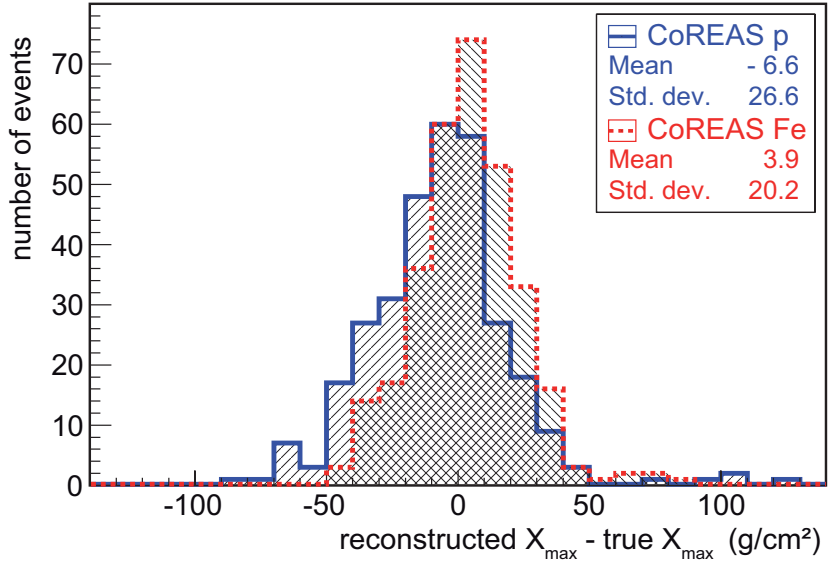}
  \caption{Deviation between the true atmospheric depths of the shower maximum, $X_\mathrm{max}$, of the simulations and the values reconstructed from the hyperbolic wavefront.}
   \label{fig_XmaxDeviation}
\end{figure}

The proportionality constant has been determined by dividing the average true $X_\mathrm{max}$ of the simulations by the average $\rho_\mathrm{cor}$. It is $25400\,$g/cm\textsuperscript{2} for the proton simulations, and $24900\,$g/cm\textsuperscript{2} for the iron simulations. For the LOPES measurements, we cannot determine this constant, since we have no detector featuring a direct measurement of $X_\mathrm{max}$. Thus, to keep the approach simple, we again use the average value of the proportionality constant, namely $25200\,$g/cm\textsuperscript{2}, to reconstruct $X_\mathrm{max}$ for all simulations and measurements. This means that a change in the cone angle $\rho_\mathrm{cor}$ of $0.001\,$rad corresponds to a change in $X_\mathrm{max}$ of approximately $25\,$g/cm\textsuperscript{2}.

For the simulations, the accuracy of the reconstructed $X_\mathrm{max}$ is not limited by the uncertainties of the individual data points, but by systematic uncertainties of the method. Thus, we have not propagated individual uncertainties to the reconstructed $X_\mathrm{max}$. Instead, we determined the systematic uncertainty of the method by comparing the reconstructed with the true $X_\mathrm{max}$ (figure \ref{fig_XmaxDeviation}). The standard deviation of the difference between the true and the reconstructed $X_\mathrm{max}$ is a measure for the systematic uncertainty of the method and is in the order of $25\,$g/cm\textsuperscript{2}. 
Moreover, there is a slight bias depending on the primary mass. The mean value of the proton and iron simulations is shifted against each other by about $10\,$g/cm\textsuperscript{2}. This means that the bias is small compared to the general systematic uncertainty and that the total combined resolution of the method is $\lesssim 30\,$g/cm\textsuperscript{2}.

\begin{figure}
  \centering
  \includegraphics[width=0.7\columnwidth]{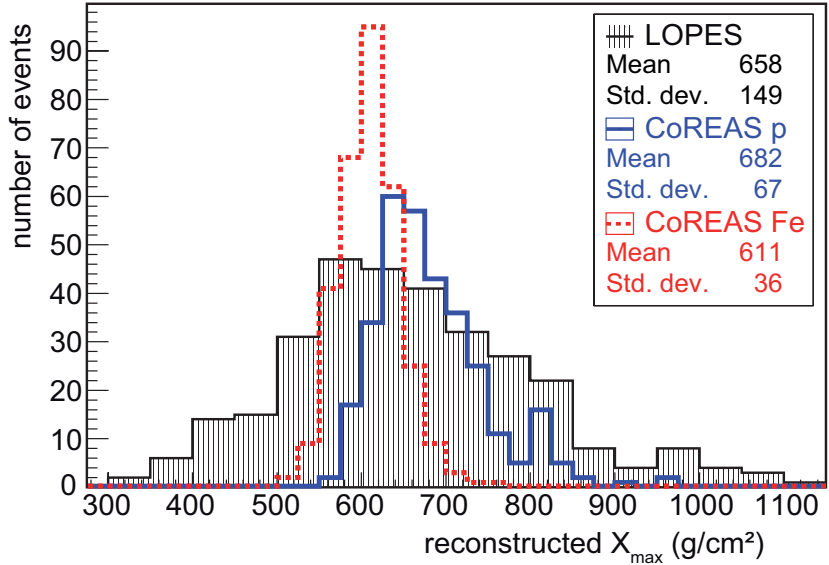}
  \caption{Reconstructed atmospheric depth of the shower maximum, $X_\mathrm{max}$. As expected from composition results of other experiments, the mean of the measured distribution is bracketed by the simulations for the extreme cases of a pure proton and a pure iron composition. The width of the measured distribution is significantly larger than the width for the simulations, which is mostly due to the larger uncertainties, and partially because the real composition is mixed.}
   \label{fig_XmaxDistributions}
\end{figure}

Finally, we have applied the reconstruction method not only to the simulations, but also to the LOPES measurements of the wavefront (figure \ref{fig_XmaxDistributions}). The mean of the measured $X_\mathrm{max}$ distribution is in between the extreme assumptions of a pure proton and a pure iron composition, as expected for simulations compatible with the measurements. Unfortunately, it is not possible to experimentally test the uncertainty of $X_\mathrm{max}$ due to the lack of an independent measurement for comparison. Moreover, we cannot determine the experimental uncertainty by error propagation, since the cross-correlation beamforming does not allow to determine the uncertainty of the individually measured cone angles.

Nevertheless, the experimental uncertainty can be estimated from the width of the histogram in figure \ref{fig_XmaxDistributions}, since the measured $X_\mathrm{max}$ distribution is a convolution of the true spread in $X_\mathrm{max}$ and the resolution. Results from Tunka indicate that the $X_\mathrm{max}$ distribution around $10^{17}\,$eV could have a width in the order of $55\,$g/cm\textsuperscript{2} \cite{EpimakhovICRC2013} which is consistent with a rough estimate based on the results shown in the reference \cite{KampertUnger2012}. Quadratically subtracting this width from the measured width, we estimate the experimental precision for the $X_\mathrm{max}$ reconstruction based on the wavefront measurements with LOPES to approximately $140\,$g/cm\textsuperscript{2}. Thus, the total uncertainty at LOPES is large compared to the systematic uncertainty of the method itself, and consequently is dominated by measurement uncertainties.

\section{Discussion}

While the simulations provide clear results in all aspects applying the presented reconstruction method, the measurements are limited by uncertainties and all observed correlations and conclusions are significantly weaker than for the simulations. Nevertheless, the measurements are compatible with the simulations, and both support the two essential results of this analysis:
\begin{itemize}
 \item The radio wavefront is sufficiently well described by a hyperboloid, but neither by a plane nor by a sphere. At larger distances a cone seems to be a sufficient approximation.
 \item The angle $\rho$ of the asymptotic cone of the hyperbola depends on the shower development and, thus, on the atmospheric depth of the shower maximum $X_\mathrm{max}$.
\end{itemize}

While we are confident in the results on a qualitative level, the exact constant and even the formulas for the dependence of $\rho$ on $X_\mathrm{max}$ and on the zenith angle $\theta$ might depend on many details, in particular: the models and codes used for the simulations, the exact conditions of the atmosphere during the development of a specific air shower, the altitude, geometric configuration and the sensitive frequency band of the antenna array. However, compared to the achieved experimental accuracy, these will all be second order corrections. Consequently, for this analysis we put the emphasis on the principle method. Nevertheless, the quantitative details have to be studied when aiming at a higher accuracy.

A better experimental accuracy can likely be achieved by experiments in locations with lower radio background. Moreover, the accuracy can be improved by denser arrays, like LOFAR \cite{CorstanjeLOFAR_wavefront2014}, due to the better sampling of the wavefront, and by larger arrays, like AERA \cite{Melissas_ARENA2012} and Tunka-Rex \cite{TunkaRexRICAP2013}, due to the larger lever arm for the determination of $\rho$. The latter two radio arrays also feature a complementary $X_\mathrm{max}$ measurement by other detectors, namely air-fluorescence and air-Cherenkov measurements, respectively, which enable a direct test of the radio measurement and a cross-calibration.

The $X_\mathrm{max}$ accuracy of better than $30\,$g/cm\textsuperscript{2} achieved for the simulations probably is not yet the theoretical limit. On the one hand, it is optimistic, since the simulations do not include uncertainties, neither for the timing nor for the geometry. On the other hand, the precision still could be improved with a more sophisticated approach, e.g., by taking into account the asymmetry of the wavefront or by using other fit functions. In addition, the $X_\mathrm{max}$ reconstruction could be based not only on the cone angle $\rho$, but also on the offset parameter $b$ of the hyperbolic wavefront. Since, $b$ and $\rho$ show some degree of correlation when letting both free in the wavefront fit, also $b$ carries some information on $X_\mathrm{max}$, which could be extracted in a more sophisticated approach.

Furthermore, the accuracy for $X_\mathrm{max}$ could be increased with multivariate analyses. In addition to the wavefront, also the frequency spectrum of the radio signal \cite{Grebe_ARENA2012} and the lateral distribution \cite{2012ApelLOPES_MTD, 2014ApelLOPES_MassComposition} depend on the shower development, and can be measured almost independently from each other. A combination of several methods has not yet been performed, but could be the next step. Since the wavefront reconstruction depends on time measurements, and the lateral distribution on amplitude measurements, both methods access complimentary information, and the combined accuracy should be better than the accuracy of each single method.

With LOPES, we already have reconstructed the shower maximum based on the lateral distribution \cite{2014ApelLOPES_MassComposition}, and obtained a mean $X_\mathrm{max}$ of $633\,$g/cm\textsuperscript{2} and an upper limit for the reconstruction precision of approximately $90\,$g/cm\textsuperscript{2}, while the mean $X_\mathrm{max}$ with the wavefront method is $658\,$g/cm\textsuperscript{2} and the reconstruction precision is approximately $140\,$g/cm\textsuperscript{2}.

The reason for the different precision could be that both methods (wavefront and lateral distribution) suffer from the noisy environment of LOPES, but the wavefront method in addition suffers significantly from uncertainties of the shower geometry. The reason for the difference in the mean $X_\mathrm{max}$ might be explained by a systematic bias due to timing uncertainties: in reference \cite{LatridouARENA2012}, it was shown for selected array geometries and positions of a radio source that in the presence of timing uncertainties, the reconstructed position of a point source is biased towards smaller distance to the radio array. Probably, this effect also means that a timing uncertainty does not just enlarge the uncertainty of the cone angle $\rho$, but also introduces a bias towards larger values of $\rho$. Still, this bias seems to be small compared to the achieved precision.

Last but not least, the accuracy for the mass composition could be enhanced by hybrid measurements, in particular the combination of radio arrays with muon detectors. Since the radio signal is emitted by the electromagnetic component of the air shower, it depends only indirectly on the muonic component. Thus, a direct measurement of the muonic component would contribute in a complementary way to the estimation of the type of the primary particle.

The slight difference in the correlation of $\rho$ and $X_\mathrm{max}$ between proton and iron induced showers probably means that also the radio signal is not totally independent of other shower parameters apart from the energy and $X_\mathrm{max}$. But this dependence seems to be small compared to the dependence on the shower maximum, and thus a complementary measurement should add significant information.

Likewise the small, but significant difference between proton and iron simulations in the zenith dependency of $\rho$ shows that the wavefront shape does not only depend on the geometric distance to the shower maximum. The shower inclination affects the distance to the shower maximum in the same way for all showers, independent of the primary particle. Thus, also here the small difference between proton and iron showers expresses some sensitivity of the radio signal to other shower parameters than just the shower maximum. The reason could be that the longitudinal development of iron showers is more compressed compared to proton showers.

Finally, the radio wavefront can be used for other purposes than just the reconstruction of the shower maximum. For example, it could be used to distinguish air showers from disturbances (RFI). While air showers have a hyperbolic wavefront, RFI sources, like air planes, can be approximated as point sources and feature a spherical wavefront.

\section{Conclusion}
We presented a first systematic study of the radio wavefront of air showers based on LOPES measurements as well as CoREAS simulations made for the situation of LOPES, in particular its altitude, geomagnetic field, and effective bandwidth. The simulated wavefront shows a slight asymmetry, which probably is due to the interference of the geomagnetic and the Askaryan radio emission. Still, compared to the measurement uncertainties, the wavefront is sufficiently well described by a symmetric hyperboloid, which can be simplified to a cone for axis distances $\gtrsim 50\,$m. The cone angle $\rho$ depends in first order on the distance of the shower maximum and can be used to reconstruct $X_\mathrm{max}$.

Following the prototype character of LOPES, we put the emphasis on outlining the principle dependencies and the potential of this method in its simplest form. Improvements are likely possible, and desirable to achieve the best accuracy for $X_\mathrm{max}$. At LOPES, however the accuracy is limited by large measurement uncertainties. Nevertheless, the CoREAS simulations indicate that the $X_\mathrm{max}$ precision could be competitive with the currently best methods, namely air-fluorescence and air-Cherenkov measurements, provided that the measurement uncertainties of the wavefront are small enough. Consequently, the prospects of the wavefront method for $X_\mathrm{max}$ reconstruction lies in its application at other radio arrays in environments with lower radio background, like AERA, Tunka-Rex and LOFAR.

The improved knowledge on the radio wavefront is beneficial also in other aspects: Applying the hyperbolic wavefront improves the reconstruction of the shower geometry, in particular the arrival direction, and potentially also the shower core. Now, in contrast to the situation a few years ago, the reconstruction is no longer limited by the missing knowledge of the correct wavefront \cite{NiglDirection2008}. Moreover, based on our measurements of the cone angle, it is possible to estimate what the impact of a simplified wavefront in the radio reconstruction would be. For example, in a plane wave reconstruction, the typical error on the arrival direction would be in the order of $\rho$, i.e., for LOPES in the order of $1^\circ$, and even larger for radio arrays at higher altitudes, since they are closer to the shower maximum. Consequently, we consider the improved description of the wavefront an important input for any future radio measurements of air showers when aiming at highest possible precision.

\acknowledgments

LOPES and KASCADE-Grande have been supported by the German Federal Ministry of Education and Research. KASCADE-Grande is partly supported by the MIUR and INAF of Italy, the Polish Ministry of Science and Higher Education and by the Romanian Authority for Scientific Research UEFISCDI (PNII-IDEI grant 271/2011). This research has been supported by grant number VH-NG-413 of the Helmholtz Association. The present study is supported by the \lq{}Helmholtz Alliance for Astroparticle Physics - HAP\rq{} funded by the Initiative and Networking Fund of the Helmholtz Association, Germany. We thank Johanna Lapp for her work on the beamforming code in the frame of her bachelor thesis.

\bibliographystyle{JHEP}
\bibliography{radioWavefrontPaper}

\providecommand{\href}[2]{#2}\begingroup\raggedright\begin{thebibliography}{10}

\bibitem{2012ApelLOPES_MTD}
{W.~D.~Apel et al.~- LOPES Collaboration}, {\it {Experimental evidence for the
  sensitivity of the air-shower radio signal to the longitudinal shower
  development}},  {\em Physical Review D} {\bf 85} (2012) 071101(R).

\bibitem{BuitinkLOFARIcrc2013}
{S.~Buitink et al.~- LOFAR Collaboration}, {\it {Shower Xmax determination
  based on LOFAR radio measurements}},  in {\em {Proceedings of the 33rd ICRC,
  Rio de Janeiro, Brazil}}, p.~\#0579, 2013.

\bibitem{RevenuExperimentsOverview_ARENA2012}
B.~{Revenu}, {\it {Overview of MHz air shower radio experiments and results}},
  {\em Proceedings of ARENA 2012 (Erlangen, Germany), AIP Conference
  Proceedings} {\bf 1535} (2013) 56--62.

\bibitem{HuegeTheoryOverview_ARENA2012}
T.~{Huege}, {\it {Theory and simulations of air shower radio emission}},  {\em
  Proceedings of ARENA 2012 (Erlangen, Germany), AIP Conference Proceedings}
  {\bf 1535} (2013) 121--127.

\bibitem{HuegeIcrc2013}
{T.~Huege}, {\it {The renaissance of radio detection of cosmic rays}},  in {\em
  {Proceedings of the 33rd ICRC, Rio de Janeiro, Brazil}}, p.~\#1294, 2013.
\newblock {arxiv.org:1310.6927}.

\bibitem{KahnLerche1966}
F.~D. {Kahn} and I.~{Lerche}, {\it {Radiation from cosmic ray air showers}},
  in {\em Proceedings of the Royal Society of London. Series A, Mathematical
  and Physical Sciences}, vol.~289, p.~206, 1966.

\bibitem{FalckeGorham2003}
H.~{Falcke} and P.~W. {Gorham}, {\it {Detecting radio emission from cosmic ray
  air showers and neutrinos with a digital radio telescope}},  {\em
  Astroparticle Physics} {\bf 19} (2003) 477--494.

\bibitem{Askaryan1962}
G.~A. {Askaryan}, {\it {Excess negative charge of an electron-photon shower and
  its coherent radio emission}},  {\em Soviet Physics JETP} {\bf 14} (1962)
  441.

\bibitem{AugerAERApolarization2014}
{Pierre Auger Collaboration}, {\it {Probing the radio emission from
  cosmic-ray-induced air showers by polarization measurements}},  {\em Physical
  Review D} {\bf 89} (2014) 052002.

\bibitem{Werner2012}
K.~{Werner}, K.~D. {de Vries}, and O.~{Scholten}, {\it {A Realistic Treatment
  of Geomagnetic Cherenkov Radiation from Cosmic Ray Air Showers}},  {\em
  Astroparticle Physics} {\bf 37} (2012) 5--16.

\bibitem{HuegeCoREAS_ARENA2012}
T.~{Huege}, M.~{Ludwig}, and C.~{James}, {\it {Simulating radio emission from
  air showers with CoREAS}},  {\em Proceedings of ARENA 2012 (Erlangen,
  Germany), AIP Conference Proceedings} {\bf 1535} (2013) 128--132.

\bibitem{SchellartLOFAR2013}
{P.~Schellart et al.~- LOFAR Collaboration}, {\it {Detecting cosmic rays with
  the LOFAR radio telescope}},  {\em Astronomy \& Astrophysics} {\bf 560}
  (2013) A98.

\bibitem{Melissas_ARENA2012}
{M.~Melissas et al.~-Pierre Auger Collaboration}, {\it {Recent Developments of
  the Auger Engineering Radio Array}},  {\em Proceedings of ARENA 2012
  (Erlangen, Germany), AIP Conference Proceedings} {\bf 1535} (2013) 63--67.

\bibitem{TunkaRexRICAP2013}
{D.~Kostunin et al.~- Tunka Collaboration}, {\it {Tunka-Rex: Status and results
  of the first measurements}},  in {\em Nucl.~Instr.~and Meth.~A; Proceedings
  of RICAP 2013, Roma, Italy}, vol.~742, pp.~89--94, 2014.

\bibitem{HuegeUlrichEngel2008}
T.~{Huege}, R.~{Ulrich}, and R.~{Engel}, {\it {Dependence of geosynchrotron
  radio emission on the energy and depth of maximum of cosmic ray showers}},
  {\em Astroparticle Physics} {\bf 30} (2008) 96--104.

\bibitem{deVries2010}
K.~D. {de Vries}, A.~M. {van den Berg}, O.~{Scholten}, and K.~{Werner}, {\it
  {The Lateral Distribution Function of Coherent Radio Emission from Extensive
  Air Showers; Determining the Chemical Composition of Cosmic Rays}},  {\em
  Astroparticle Physics} {\bf 34} (2010) 267--273.

\bibitem{2014ApelLOPES_MassComposition}
{W.~D.~Apel et al.~- LOPES Collaboration}, {\it {Reconstruction of the energy
  and depth of maximum of cosmic-ray air-showers from LOPES radio
  measurements}},  {\em Physical Review D} (2014). submitted, under review.

\bibitem{Grebe_ARENA2012}
{S.~Grebe et al.~-Pierre Auger Collaboration}, {\it {Spectral index analysis of
  the data from the Auger Engineering Radio Array}},  {\em Proceedings of ARENA
  2012 (Erlangen, Germany), AIP Conference Proceedings} {\bf 1535} (2013)
  73--77.

\bibitem{Lafebre2010}
S.~{Lafebre}, H.~{Falcke}, J.~{H\"orandel}, et~al., {\it {Prospects for
  determining air shower characteristics through geosynchrotron emission
  arrival times}},  {\em Astroparticle Physics} {\bf 34} (2010) 12--17.

\bibitem{CODALEMAMarinICRC2011}
{V.~Marin et al.~- CODALEMA Collaboration}, {\it {Charge excess signature in
  the CODALEMA data. Interpretation with SELFAS2}},  in {\em {Proceedings of
  the 32nd ICRC, Beijing, China}}, vol.~1, p.~\#0942, 2011.
\newblock {www.ihep.ac.cn/english/conference/icrc2011/paper}.

\bibitem{NellesLOFAR_LDF2014}
A.~{Nelles}, S.~{Buitink}, H.~{Falcke}, et~al., {\it {A parameterization for
  the radio emission of air showers as predicted by CoREAS simulations and
  applied to LOFAR measurements}},  {\em Astroparticle Physics} {\bf 60} (2014)
  13--24.

\bibitem{NiglDirection2008}
{A.~Nigl et al.~- LOPES Collaboration}, {\it {Direction identification in radio
  images of cosmic-ray air showers detected with LOPES and KASCADE}},  {\em
  Astronomy \& Astrophysics} {\bf 487} (2008) 781--788.

\bibitem{CorstanjeLOFAR_wavefront2014}
{A.~Corstanje et al.~-LOFAR Collaboration}, {\it {The shape of the radio
  wavefront of extensive air showers as measured with LOFAR}},  {\em
  Astroparticle Physics} (2014). {DOI:10.1016/j.astropartphys.2014.06.001, in
  press}.

\bibitem{AntoniApelBadea2003}
{T.~Antoni et al.~- KASCADE Collaboration}, {\it {The Cosmic-Ray Experiment
  KASCADE}},  {\em Nuclear Instruments and Methods in Physics Research A} {\bf
  513} (2003) 490--510.

\bibitem{FalckeNature2005}
{H.~Falcke et al.~- LOPES Collaboration}, {\it {Detection and imaging of
  atmospheric radio flashes from cosmic ray air showers}},  {\em Nature} {\bf
  435} (2005) 313--316.

\bibitem{HuegeARENA_LOPESSummary2010}
{T.~Huege et al.~- LOPES Collaboration}, {\it {The LOPES experiment - recent
  results, status and perspectives}},  in {\em Nucl.~Instr.~and Meth.~A;
  Proceedings of the ARENA 2010 conference, Nantes, France}, vol.~662,
  Supplement 1, pp.~S72--S79, 2012.

\bibitem{SchroederLOPESsummaryARENA2012}
{F.~G.~Schr\"oder et al.~- LOPES Collaboration}, {\it {Cosmic Ray Measurements
  with LOPES: Status and Recent Results}},  in {\em AIP Conf.~Proc.~1535},
  vol.~1535, pp.~78--83, 2012.
\newblock Proc.~5th ARENA, Erlangen, Germany.

\bibitem{2013ApelLOPESlateralComparison}
{W.~D.~Apel et al.~- LOPES Collaboration}, {\it {Comparing LOPES measurements
  of air-shower radio emission with REAS 3.11 and CoREAS simulations}},  {\em
  Astroparticle Physics} {\bf 50-52} (2013) 76--91.

\bibitem{HeckKnappCapdevielle1998}
D.~{Heck}, J.~{Knapp}, J.~N. {Capdevielle}, et~al., {\it {CORSIKA: A Monte
  Carlo Code to Simulate Extensive Air Showers}},  FZKA Report 6019,
  Forschungszentrum Karlsruhe, 1998.

\bibitem{JamesEndPoint2010}
C.~{James}, H.~{Falcke}, T.~{Huege}, and M.~{Ludwig}, {\it {General description
  of electromagnetic radiation processes based on instantaneous particle
  acceleration in 'endpoints'}},  {\em Physical Review E} {\bf 84} (2011)
  056602.

\bibitem{OstapchenkoQGSjetII2006}
S.~{Ostapchenko}, {\it {QGSJET-II: results for extensive air showers}},  {\em
  Nuclear Physics B - Proceedings Supplements} {\bf 151} (2006) 147--150.

\bibitem{BattistoniFLUKA2007}
G.~{Battistoni} et~al., {\it {The FLUKA code: Description and benchmarking}},
  {\em Proceedings of the Hadronic Shower Simulation Workshop 2006, AIP
  Conference Proceeding} {\bf 896} (2007) 31--49.

\bibitem{LudwigREAS3_2010}
M.~{Ludwig} and T.~{Huege}, {\it {REAS3: Monte Carlo simulations of radio
  emission from cosmic ray air showers using an \lq end-point\rq~formalism}},
  {\em Astroparticle Physics} {\bf 34} (2011) 438--446.

\bibitem{CRtools}
``{Cosmic Ray Tools - The Analysis Software of LOPES}.''
\newblock
  {http://usg.lofar.org/svn/code/branches/cr-tools-stable/src/CR-Tools/apps/ca%
ll\_pipeline.cc}.

\bibitem{SchroederTimeCalibration2010}
F.~G. {Schr\"oder}, T.~{Asch}, L.~{B\"ahren}, et~al., {\it {New method for the
  time calibration of an interferometric radio antenna array}},  {\em Nuclear
  Instruments and Methods in Physics Research A} {\bf 615} (2010) 277--284.

\bibitem{SchroederNoise2010}
{F.~G.~Schr\"oder et al.~- LOPES Collaboration}, {\it {On noise treatment in
  radio measurements of cosmic ray air showers}},  in {\em Nucl.~Instr.~and
  Meth.~A; Proceedings of the ARENA 2010 conference, Nantes, France}, vol.~662,
  Supplement 1, pp.~S238--S241, 2012.

\bibitem{2010ApelLOPESlateral}
{W.~D.~Apel et al.~- LOPES Collaboration}, {\it {Lateral distribution of the
  radio signal in extensive air showers measured with LOPES}},  {\em
  Astroparticle Physics} {\bf 32} (2010) 294--303.

\bibitem{EpimakhovICRC2013}
{S.~Epimakhov et al.~- Tunka-133 Collaboration}, {\it {Elemental Composition of
  Cosmic Rays above the Knee from Xmax measurements of the Tunka Array}},  in
  {\em {Proceedings of the 33rd ICRC, Rio de Janeiro, Brazil}}, p.~\#0326,
  2013.

\bibitem{KampertUnger2012}
K.-H. {Kampert} and M.~{Unger}, {\it {Measurements of the cosmic ray
  composition with air shower experiments}},  {\em Astroparticle Physics} {\bf
  35} (2012) 660--678.

\bibitem{LatridouARENA2012}
P.~{Lautridou}, O.~{Ravel}, A.~{Rebai}, and A.~{Lecacheux}, {\it {Some possible
  interpretations from data of the CODALEMA experiment}},  in {\em AIP
  Conf.~Proc.~1535}, vol.~1535, pp.~99--104, 2012.
\newblock Proc.~5th ARENA, Erlangen, Germany.

\end{thebibliography}\endgroup

\end{document}